\documentclass[fleqn,10pt]{wlscirep}
\usepackage[utf8]{inputenc}
\usepackage[T1]{fontenc}
\usepackage{graphicx}
\usepackage{booktabs}
\usepackage{adjustbox}
\usepackage{subfigure}
\usepackage{float}
\title{Federated Learning-based Semantic Segmentation for Lane and Object Detection in Autonomous Driving}

\author[1,*]{Gharbi Khamis Alshammari}
\author[2,*]{Ahmad Abubakar}
\author[3]{Nada M.O. Sid Ahmed}
\author[4]{Naif Khalaf Alshammari}
\affil[1]{Department of Artificial Intelligence, College of Computer Science and Engineering, University of Ha’il, Hail 8145, Saudi Arabia}
\affil[2]{Center for Autonomous Robotic Systems (KUCARS), Khalifa University, Abu Dhabi, United Arab Emirates.}
\affil[3]{Department of Information and Computer Science, College of Computer Science and Engineering, University of Ha’il, Hail 55211, Saudi Arabia}
\affil[4]{Mechanical Engineering Department, Engineering College, University of Ha’il, Hail 8148, Saudi Arabia}

\affil[*]{gk.alshammari@uoh.edu.sa and 100059792@ku.ac.ae}

\begin{abstract}
Autonomous Vehicles (AVs) require precise lane and object detection to ensure safe navigation. However, centralized deep learning (DL) approaches for semantic segmentation raise privacy and scalability challenges, particularly when handling sensitive data. This research presents a new federated learning (FL) framework that integrates secure deep Convolutional Neural Networks (CNNs) and Differential Privacy (DP) to address these issues. The core contribution of this work involves: (1) developing a new hybrid UNet-ResNet34 architecture for centralized semantic segmentation to achieve high accuracy and Mean Intersection over Union (Mean IoU) that tackles privacy concerns due to centralized training (2) Implementing the privacy-preserving FL model, distributed across AVs to enhance performance through secure CNNs and DP mechanisms. In the proposed FL framework, the methodology systematically distinguishes itself from the existing approach through the following: (i) ensuring data decentralization through FL to uphold user privacy by eliminating the need for centralized data aggregation; (ii) integrating DP mechanisms to secure sensitive model updates against potential adversarial inference attacks; and (iii) evaluating the framework’s performance and generalizability using RGB and semantic segmentation (SEG) datasets derived from the CARLA simulator. Experimental results show significant improvements in accuracy, from 81.5\% to 88.7\% for the RGB dataset and from 79.3\% to 86.9\% for the SEG dataset over 20 to 70 Communication Rounds (CRs). Global loss was reduced by over 60\%, and minor accuracy trade-offs from DP were observed. This study contributes by offering a scalable, privacy-preserving FL framework tailored for AVs, optimizing communication efficiency while balancing performance and data security.
\end{abstract}
\begin{document}

\flushbottom
\maketitle
%
%
\thispagestyle{empty}

\noindent Please note: Abbreviations should be introduced at the first mention in the main text – no abbreviations lists. Suggested structure of main text (not enforced) is provided below.

\section*{Introduction}

AVs rely on accurate lane and object detection to ensure safe navigation in complex driving environments~\cite{10.1007/978-981-16-2406-3_70}.
Semantic segmentation, which assigns labels to every pixel in an image, is a crucial task for AV perception systems. However, real-world scenarios present challenges due to variations in road conditions, lighting, weather, and occlusions, which make achieving high segmentation accuracy difficult~\cite{10649172,bathla2022autonomous,ZHANG2023103458,9016391}. Current centralized DL approaches for semantic segmentation require large amounts of sensitive driving data to be stored on central servers, raising privacy concerns and potential regulatory violations\cite{a17030103, tang2023surveyautomateddrivingtesting,electronics12051199}.
While centralized models like UNet, ResNet, and SegNet have shown strong performance for lane and object detection, they present significant drawbacks. Privacy concerns arise when sensitive data is shared and stored centrally\cite{Dunphy2024Ensuring}. These methods also require substantial computational resources, limiting their scalability in real-world AV environments. Furthermore, existing models struggle with non-Independent and Identically Distributed (non-IID) data, where each AV captures data under different conditions, causing inconsistencies in model performance\cite{10564103}.
Addressing these challenges requires a decentralized solution capable of maintaining privacy, optimizing communication efficiency, and accommodating the computational limitations of AVs. Our approach, FL, offers a promising direction by enabling model training across multiple AVs without sharing raw data. However, implementing secure FL for semantic segmentation comes with its own
challenges, including handling non-IID data, ensuring robust privacy through Differential Privacy (DP), and maintaining communication efficiency\cite{ALHUTHAIFI2023833}.
Several DL architectures have been proposed for semantic segmentation in AVs, such as UNet, ResNet34, and SegNet, which rely on centralized training\cite{10294507}. Although these methods achieve high accuracy in controlled environments, their centralized nature introduces substantial privacy risks and poses significant challenges in terms of computational demand. The aggregation of large datasets on a central server raises security issues, particularly under strict privacy regulations like GDPR and CCPA\cite{amonkar2023comparative,Mayeke2024}. Additionally, these architectures perform poorly with non-IID data, commonly observed in AV scenarios, leading to suboptimal convergence and degraded performance.
In this study, we propose an enhanced FL framework that integrates secure deep CNNs and Differential Privacy (DP) techniques to address these challenges. 
The key claims of our proposed framework include: (i) Data decentralization via FL: FL allows AVs to collaboratively train models without sharing raw data, significantly reducing privacy risks. (ii) Differential Privacy integration: DP ensures that model updates are protected, further safeguarding sensitive information while maintaining robust performance. (iii) Hybrid architecture: We combine UNet’s segmentation capabilities with ResNet34’s feature extraction strengths, creating a hybrid model for accurate lane and object detection. (iv) Communication efficiency: Optimized communication protocols are implemented to reduce bandwidth usage and latency, essential for real-time AV environments.

This research aims to improve semantic segmentation in AVs by addressing the aforementioned challenges. The main contributions of this research are as follows:

\begin{enumerate}
     
\item Hybrid UNet-ResNet34 for FL-Based Semantic Segmentation: The proposed hybrid UNet-ResNet34 model addresses the lack of adaptability in standard segmentation models (e.g., UNet, ResNet34) for non-IID federated learning scenarios by combining UNet’s fine-grained segmentation capabilities with ResNet34’s deep feature extraction, achieving superior performance in diverse lighting and environmental conditions on AV datasets compared to conventional architectures.
    
\item Privacy-Preserving Federated Learning (FL) Framework: Our FL framework eliminates data-sharing risks by training models directly on autonomous vehicles (AVs) instead of relying on centralized data aggregation, ensures decentralized learning while maintaining segmentation accuracy across heterogeneous AV datasets, and achieves comparable performance with significantly reduced privacy risks, as demonstrated by experimental results.
    
\item Differential Privacy (DP) for Secure Model Updates: To further enhance security, we integrate Differential Privacy (DP) mechanisms into Federated Learning (FL) to mitigate adversarial inference risks by adding controlled noise to model updates, ensuring privacy while maintaining high segmentation performance, with evaluations demonstrating that our DP-FL approach retains over 90\% of the original accuracy alongside strong privacy guarantees.
    
\item Communication-Efficient FL for Scalable AV Networks: Federated Learning (FL) introduces communication overhead due to repeated model updates, but by optimizing communication protocols to reduce bandwidth consumption without sacrificing accuracy, our results demonstrate that increasing communication rounds (CRs) from 20 to 70 improves accuracy from 81.5\% to 88.7\% on the RGB dataset and from 79.3\% to 86.9\% on the SEG dataset, achieving these enhancements without excessive communication costs and thereby making FL feasible for real-world autonomous vehicle (AV) applications. 
    
\item Robust Evaluation Across AV Datasets: Our model, benchmarked on RGB and SEG datasets from the CARLA simulator, a widely used AV perception dataset, demonstrates robustness through performance metrics (Mean IoU, sensitivity, specificity, F1-score) and outperforms baseline UNet, ResNet34, and SegNet models in both centralized and federated learning settings.
 
\end{enumerate}

Our proposed approach is distinct with the existing approaches through several points, including: (a) Eliminates centralized data storage while maintaining segmentation accuracy. (b) Ensures strong privacy protection using FL and DP techniques. (c) Optimizes communication efficiency, making FL viable for AV networks. (d) Scales effectively across non-IID AV datasets, outperforming standard FL and DL models.

The remainder of this paper is organized as follows: Section 2 (Literature Review) discusses existing semantic segmentation methods, federated learning frameworks, and privacy-preserving techniques while highlighting their limitations; Section 3 (Materials and Methods) presents details on dataset selection, preprocessing, model architecture, the FL framework, and differential privacy integration; Section 4 (Experimental Setup) describes the experimental setup, implementation of the proposed FL framework,  and evaluation metrics. Section 5 (Discussion) analyzes the results, emphasizing performance improvements, trade-offs, practical implications, and comparisons with state-of-the-art approaches; and Section 6 (Conclusion and Future Work) summarizes key findings and suggests directions for future research.

\section*{LITERATURE REVIEW}

Semantic segmentation is crucial in AVs for detailed environmental understanding. DL techniques, particularly CNNs, dominate this domain, enabling pixel-wise classification of images captured by AV cameras\cite{app12146831,9684905}. Popular architectures like UNet, ResNet, and SegNet are widely used due to their capability in capturing spatial hierarchies and contextual information\cite{DBLP:journals/corr/abs-1901-06032,DBLP:journals/corr/abs-2001-05566}. However, centralized training of these models requires vast amounts of annotated data, collected from diverse locations and conditions, creating significant challenges in data aggregation and processing\cite{9284628,Liang2022,ZHANG2023103458}. Privacy concerns further arise due to the centralized storage and transfer of sensitive data\cite{KumarTyagi2020}. Additionally, training sophisticated CNNs on large datasets demands high-performance hardware, which may not be feasible for all stakeholders involved in AV operations. Moreover, data heterogeneity across different vehicles can lead to model biases, reducing the generalizability of the trained models. FL addresses privacy and computational inefficiencies inherent in centralized learning systems\cite{DBLP:journals/corr/abs-1907-09693,MOTHUKURI2021619}. It enables collaborative model training across multiple decentralized devices, each holding its local data. In FL, only model updates are shared, thus preserving data privacy and reducing the risk of breaches. FL has shown its effectiveness across applications such as healthcare, finance, and autonomous driving\cite{electronics11040670,9153560}. For AVs, FL allows vehicles to learn from shared experiences while maintaining individual privacy\cite{chellapandi2023federatedlearningconnectedautomated,9457207}. Techniques like federated averaging (FedAvg) efficiently aggregate local model updates\cite{DBLP:journals/corr/abs-2104-11375,DBLP:journals/corr/abs-2012-06706}. Nevertheless, challenges remain in autonomous driving applications, including communication overhead and non-IID data issues\cite{9016391,10054499}. Research has focused on optimizing communication protocols and ensuring robust training despite data heterogeneity\cite{6d5642637cab4ea8b478671afd9a0c2a,9153560}. Deep CNNs are fundamental in image segmentation tasks, providing powerful feature extraction through hierarchical learning\cite{DBLP:journals/corr/abs-1907-09693,Dhillon2020}. Architectures like UNet, ResNet, and the hybrid UNet-ResNet34 have achieved high accuracy in delineating object boundaries and lane markings in driving scenarios\cite{10294507}. These architectures generally employ an encoder-decoder structure: the encoder captures spatial features at multiple scales, while the decoder reconstructs the segmented image. Techniques like skip connections and dilated convolutions enhance feature propagation and capture contextual information\cite{Tiwari2023,Xu2024}. 

Mathematically, a CNN can be represented as in Eq. 1, where x is the input image, y is the segmented output, and W represents the learnable network weights. The training process minimizes a loss function, as depicted in Eq. 2, where y$*$ denotes the ground truth segmentation.

DP is a framework designed to provide formal privacy guarantees when sharing statistical information\cite{liu2021differentialprivacyrobuststatistics,s20247030}. In FL, DP adds noise to model updates, ensuring that the presence or absence of any single data point does not significantly affect outcomes\cite{MOHAMMADI2024104918,electronics11040670}. This approach mitigates the risk of re-identification attacks. DP is implemented by adding noise to gradients or weights before sharing them with the central server\cite{9714350}. The noise is typically sampled from Gaussian or Laplacian distributions, characterized by parameters $\epsilon$ (privacy budget) and $\delta$ (failure probability). Recent works in privacy-preserving FL have explored solutions to mitigate these effects. Yao et al. (2025)\cite{YAO2025107706, 4f49fba2c01a43ca89d3b37bd02f400e} introduced FedShufde, a privacy-preserving FL framework for UAV delivery systems that leverages shuffling techniques to reduce accuracy degradation caused by DP. Similarly, Zhang et al. (2024)\cite{drones8070312} proposed Fed4UL, a cloud-edge-end collaborative FL approach that improves model convergence despite non-IID data. These methods can be adapted to AV applications to maintain segmentation accuracy while ensuring data privacy.The goal is to balance model accuracy and privacy protection. One of the major challenges of integrating DP into FL is the increased computational and communication overhead, which can strain AV resources. Edge-based learning and optimized communication protocols have been proposed to address these challenges. Yao et al. (2024)\cite{YAO2024103532} developed a privacy-preserving location data collection framework for intelligent systems in edge computing, which optimizes communication strategies to minimize bandwidth usage while preserving privacy. Additionally, Dong et al. (2025)\cite{30bc600b8fd94f3e81f8d1b6d67adc3d} introduced a blockchain-aided task distribution framework for UAV communications, efficiently managing computational resources to reduce overhead. Applying such methods to AVs can help alleviate the increased computational burden caused by FL+DP, making it more scalable for real-time applications. Robustness across varying driving conditions (e.g., night, fog, rain) remains a critical area of improvement for AV segmentation models. UAV-based research has demonstrated the effectiveness of secure and adaptive FL strategies in maintaining model stability across dynamic environments. Yao et al. (2021)\cite{77f00bb0ef75473794373563af3e86af} proposed a security framework for edge-based UAV delivery systems that enhances learning robustness under different environmental conditions. Similarly, Dong et al. (2021)\cite{30bc600b8fd94f3e81f8d1b6d67adc3d} introduced a blockchain-aided self-sovereign identity framework that improves security and stability in real-time UAV applications. 
A differentially private mechanism M satisfies Eq. 3 for any two adjacent datasets D and D$'$ differing by a single element, and any output y.
\begin{equation}
    y = f(x; \mathbf{W})
\end{equation}
\begin{equation}
    \mathcal{L}(y, \mathbf{y}^*)
\end{equation}
\begin{equation}
    \mathbb{P}[\mathcal{M}(\mathcal{D}) = y] \leq e^\epsilon \mathbb{P}[\mathcal{M}(\mathcal{D}') = y] + \delta
\end{equation}

Lane detection is essential in autonomous driving systems, supporting vehicle navigation and safety. Recent advancements in DL have significantly improved lane detection accuracy and robustness. We review several state-of-the-art methods, focusing on their approaches, strengths, and limitations. LaneNet utilizes an encoder-decoder architecture for pixel-wise lane segmentation\cite{TANG2021107623}. The encoder extracts features from the input image, while the decoder generates segmentation masks. However, LaneNet faces challenges with complex road scenarios and varying lighting conditions. SCNN (Spatial CNN) introduces spatial convolutional layers to enhance continuous lane marking detection\cite{DBLP:journals/corr/abs-2110-04079}, but its high computational demands limit real-time applications. DeepLabV3+ uses atrous convolution and spatial pyramid pooling to capture multi-scale contextual information, improving segmentation accuracy\cite{WANG2022104969,10.1117/1.JEI.31.5.053006}. However, its computational intensity necessitates optimization for real-time deployment. ENet offers a lightweight architecture for real-time segmentation, balancing accuracy and efficiency\cite{Yi2023}, but it struggles with thin and continuous lane markings compared to more complex models. Several approaches have been developed to improve object detection in AVs, particularly under challenging conditions like low visibility or adverse weather. For instance, the PVDM-YOLOv8l model integrates techniques to enhance pedestrian and vehicle detection even in adverse weather, providing a reliable solution for real-time AV navigation~\cite{Tahir2024}. However, these methods rely heavily on centralized DL models, which introduce privacy concerns when sensitive AV data is aggregated on central servers.

Lane detection is essential in autonomous driving systems, supporting vehicle navigation and safety~\cite{10.1007/978-981-16-2406-3_70,Abubakar2019}. Recent advancements in DL have significantly improved lane detection accuracy and robustness. We review several state-of-the-art methods, focusing on their approaches, strengths, and limitations. LaneNet utilizes an encoder-decoder architecture for pixel-wise lane segmentation\cite{TANG2021107623}. The encoder extracts features from the input image, while the decoder generates segmentation masks. However, LaneNet faces challenges with complex road scenarios and varying lighting conditions~\cite{10649172,Abubakar2019}. SCNN (Spatial CNN) introduces spatial convolutional layers to enhance continuous lane marking detection\cite{DBLP:journals/corr/abs-2110-04079}, but its high computational demands limit real-time applications. DeepLabV3+ uses atrous convolution and spatial pyramid pooling to capture multi-scale contextual information, improving segmentation accuracy\cite{WANG2022104969,10.1117/1.JEI.31.5.053006}. However, its computational intensity necessitates optimization for real-time deployment. ENet offers a lightweight architecture for real-time segmentation, balancing accuracy and efficiency\cite{Yi2023}, but it struggles with thin and continuous lane markings compared to more complex models. Several approaches have been developed to improve object detection in AVs, particularly under challenging conditions like low visibility or adverse weather. For instance, the PVDM-YOLOv8l model integrates techniques to enhance pedestrian and vehicle detection even in adverse weather, providing a reliable solution for real-time AV navigation\cite{Tahir2024,Abubakar2019}. However, these methods rely heavily on centralized DL models, which introduce privacy concerns when sensitive AV data is aggregated on central servers.

The current state-of-the-art methods for lane detection, despite their advancements, face several limitations, including challenges in accurately detecting lanes in complex road scenarios like intersections and varied road textures, high computational demands that hinder real-time deployment on resource-limited platforms, underperformance under varying environmental conditions such as lighting, weather, and occlusions, and issues with scalability and data privacy due to centralized training methods. To address these limitations, we propose an enhanced FL framework that integrates secure deep CNNs with DP. This approach offers improved segmentation accuracy in complex road scenarios through a hybrid CNN architecture, reduces computational demands and ensures scalability by distributing the training process across clients, and addresses privacy concerns inherent in centralized methods through the integration of DP. A detailed comparison of our proposed method with state-of-the-art methods, presented in Section 4, highlights significant improvements in accuracy, computational efficiency, and robustness across various conditions.

\section*{Materials and Methods}

This study utilizes RGB and SEG datasets for semantic segmentation in AVs. The RGB dataset comprises full-color images captured by AV cameras, providing richly annotated scenes with diverse road conditions and environments. The SEG dataset contains pixel-wise labeled images for accurate segmentation of lanes, vehicles, pedestrians, and other critical objects in autonomous driving. These datasets were chosen to evaluate and enhance the FL framework’s robustness and generalization in real-world conditions. The study is organized into two phases. Phase-1 involves establishing a baseline using a centralized hybrid UNet-ResNet34 model, trained on the RGB and SEG datasets to assess initial capabilities and limitations. Key performance metrics such as accuracy, global loss, and segmentation quality were evaluated. Phase-2 extends the framework by incorporating FL, allowing decentralized model training across multiple AVs while integrating secure deep CNNs and DP to address data privacy concerns. Models were evaluated over CRs (20, 40, and 70) to study iterative learning and communication efficiency. The entire framework of the study is depicted in Figure 1.

\begin{figure}[H]
    \centering
    \includegraphics[width=0.7\linewidth]{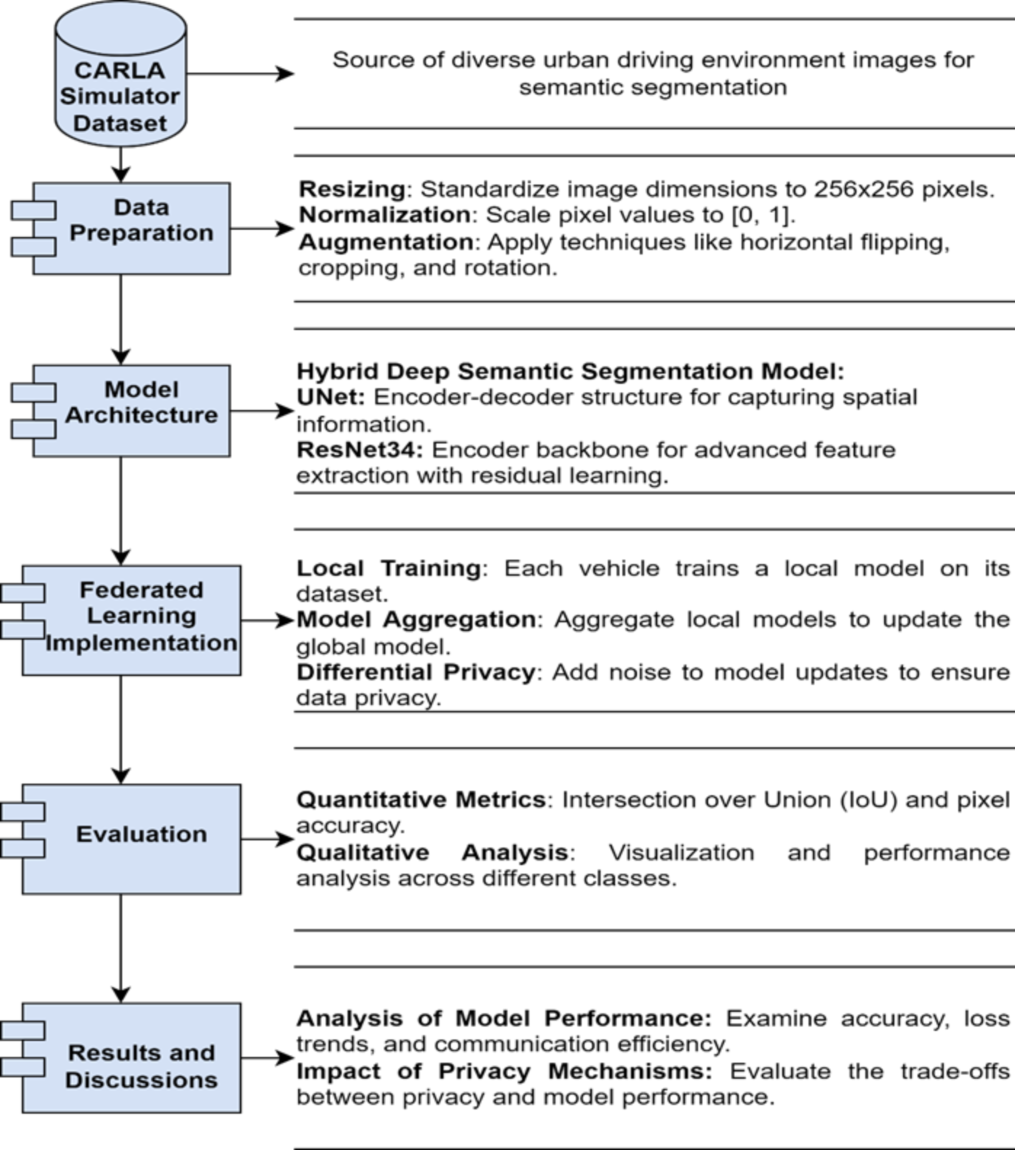}
    \caption{Overall Framework of the Proposed Research Study}
    \label{fig:enter-label}
\end{figure}

\subsection*{Data Pre-processing and Model Architecture}

The methodology begins with data preprocessing, including normalization, augmentation, and edge detection. The datasets are partitioned across multiple clients, representing decentralized AVs in an FL environment. 

\subsubsection*{Phase-1}
The hybrid UNet-ResNet34 model combines the spatial context capabilities of UNet with the residual learning of ResNet34, optimized for detecting lanes and objects in complex scenes. The sub-figures A and B of figure 2 illustrate the workflow and architecture used in Phase-1. The hybrid model processes input images through encoder-decoder layers, enhanced by residual connections and pooling layers for feature extraction. The model is trained to minimize the categorical cross-entropy loss calculated using Eq. 4. Furthermore, RGB images are normalized to the range [0, 1], resized to 256x256 pixels, and undergo data augmentation techniques such as rotation, scaling, and flipping to improve generalization. Performance Metrics: The model’s performance is assessed using accuracy, Mean IoU, sensitivity, specificity, and precision. These metrics provide a comprehensive evaluation of the model’s ability to segment lanes and objects in various driving scenarios.
\begin{equation}
    \mathcal{L} = -\sum_{i=1}^N y_i \log(\hat{y}_i)
\end{equation}
\subsubsection*{Phase-2:}
Phase-2 introduces FL for collaborative training across multiple clients, preserving data privacy through decentralized learning. The federated server aggregates model updates from clients using a weighted average, as shown in Eq. 5 and illustrated in sub-figure C of Figure 2. Experiments with FL-CNN models at 20, 40, and 70 CRs were conducted to identify the optimal balance between performance and communication efficiency. DP mechanisms were integrated by adding Gaussian noise to the model updates, as described in Eq. 6, to enhance privacy. The hybrid UNet-ResNet34 architecture was adapted for FL with DP to handle non-IID data distributions typical in AV environments, optimizing communication and ensuring scalability.
\begin{figure}[H]
    \centering
    \includegraphics[width=1.0\linewidth]{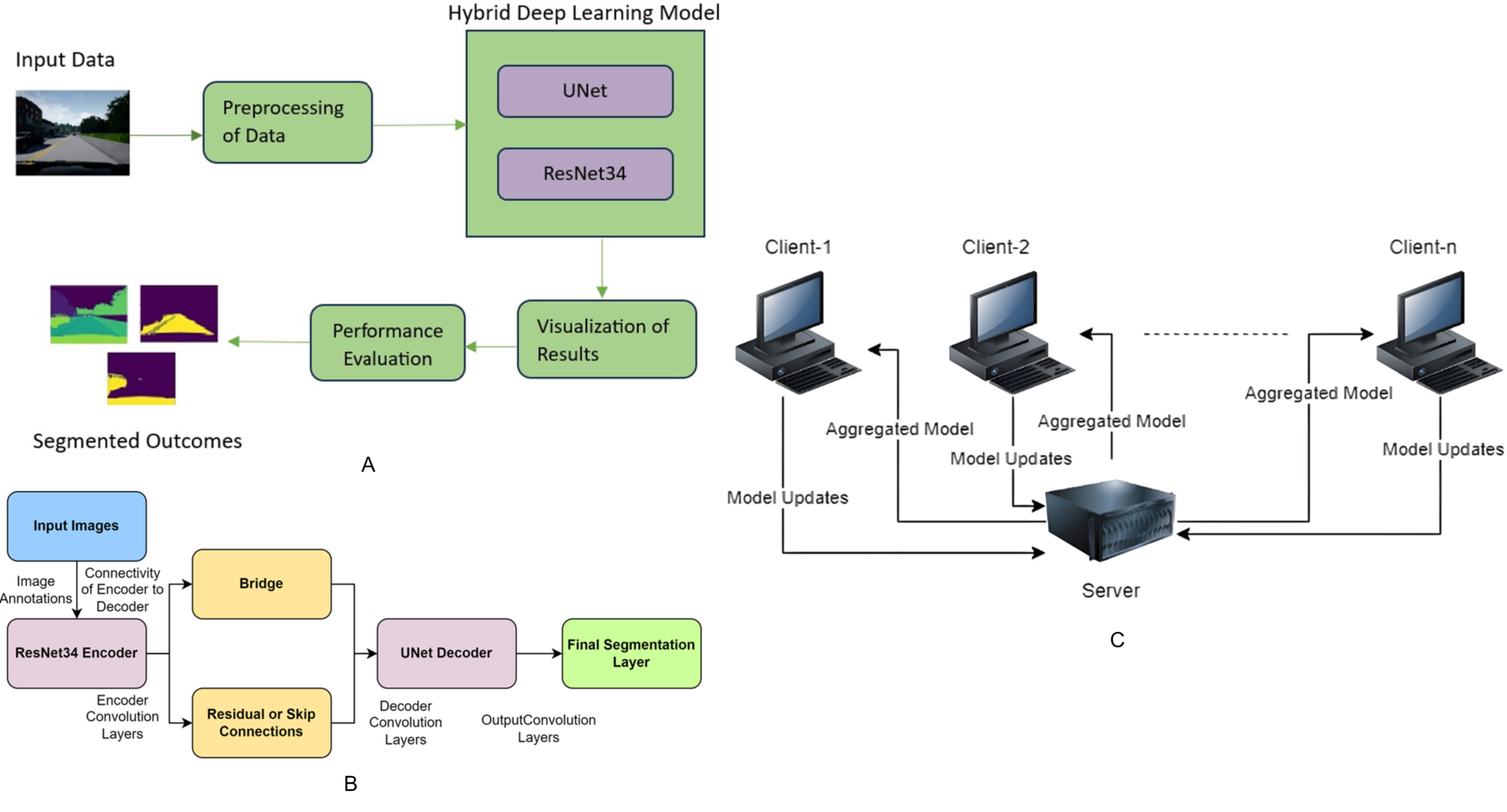}
    \caption{Phase-1 and Phase-2 Model Architectures. (A) Phase-1: Hybrid UNet-ResNet34 Framework, (B) Hybrid UNet-ResNet34 Architecture, (C) FL Workflow}
    \label{fig:enter-label}
\end{figure}
\begin{equation}
    w^{t+1} = \sum_{i=1}^N \frac{n_i}{n} w_i^{t+1}
\end{equation}
\begin{equation}
    \tilde{w}_i^{t+1} = w_i^{t+1} + \mathcal{N}(0, \sigma^2)
\end{equation}

\subsection*{Implementation}

The research uses TensorFlow, Keras, TensorFlow Federated, and TensorFlow Privacy for model development, FL orchestration, and DP implementation. Experiments are conducted on GPUs and simulated AV clients with varying CR configurations. The integrated approach employed in this study ensures a secure, scalable, and efficient FL framework for semantic segmentation in AVs, addressing key challenges in data privacy, robustness, and scalability.

\section*{Experiment Design and Setup}

This section details the experimentation process and results from Phase-1 and Phase-2 of the research. The primary focus is evaluating model performance metrics to understand their effectiveness in semantic segmentation for AVs.

The input data was sourced from the Lyft Udacity Challenge dataset\cite{Kaggle}, and additional images were captured using the CARLA driving simulator. These datasets simulate diverse and realistic urban driving environments, making them suitable for testing AV perception systems.

The RGB data set contains standard RGB images that capture a wide range of urban driving scenarios, including varied environmental conditions such as different times of day, lighting variations, and weather changes (e.g., sunny, rainy and foggy conditions). These conditions mimic the challenges AVs face in real-world settings, where accurate detection of objects like roads, vehicles, and pedestrians is critical under changing environmental factors.

The SEG dataset provides pixel-level segmentation labels from the CARLA simulator, offering ground truth annotations for semantic segmentation tasks. This dataset also reflects real-world complexity, capturing occluded objects, cluttered scenes, and varied object types (e.g., vehicles, pedestrians, road markings), which present significant challenges for accurate segmentation.
Both datasets present specific challenges for the models:
\begin{itemize}
\item Environmental Variability: The RGB dataset includes a variety of lighting and weather conditions, which test the model’s ability to generalize across different environments.
\item Object Complexity and Occlusions: The SEG dataset includes objects of varying sizes and shapes that may appear partially occluded or in dense urban settings, mimicking real-world driving scenarios where AVs must quickly and accurately identify objects in dynamic environments.
\end{itemize}

\noindent Pre-processing techniques, including image normalization using the standard scalar technique and augmentation (e.g., flipping and resizing), were applied to enhance model training. Figures 3 and Table 1 illustrate examples of images and corresponding segmentation labels from the CARLA dataset, along with the distribution of object classes in the RGB and SEG datasets, respectively, by also highlighting the uniform class distribution crucial for balanced training and accurate segmentation performance.

\begin{figure}[H]
    \centering
    \includegraphics[width=0.5\linewidth]{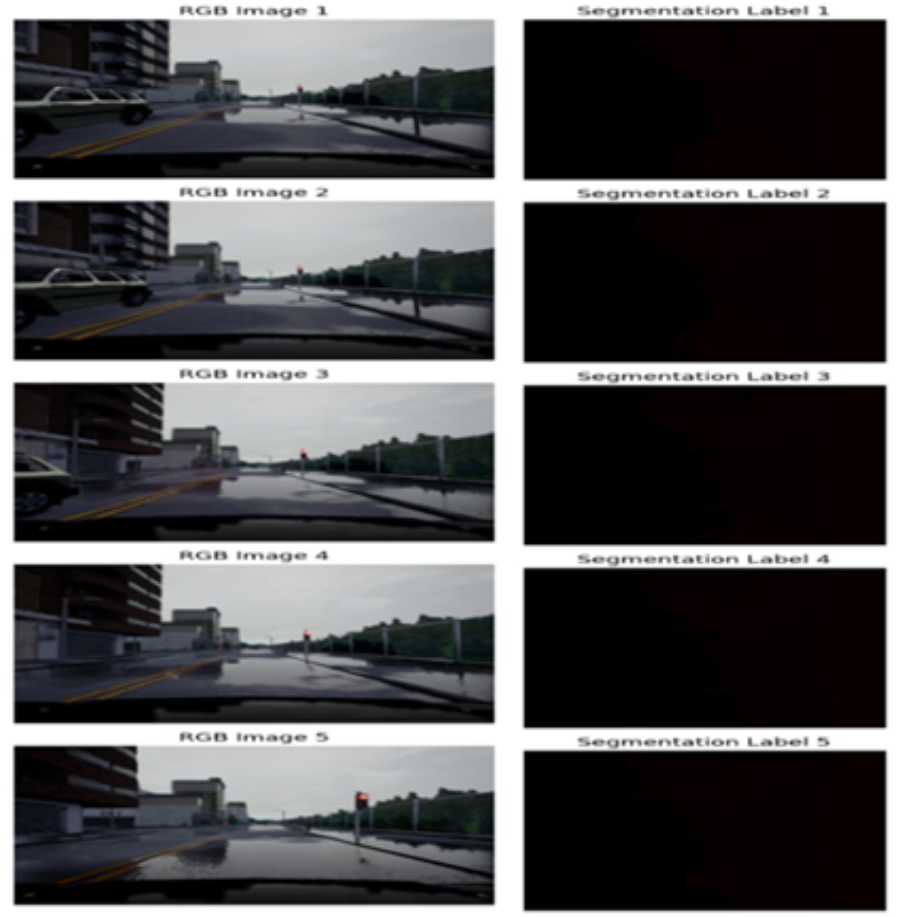}
    \caption{Examples of Images and Corresponding Segmentation Labels from the CARLA Simulator Dataset}
    \label{fig:enter-label}
\end{figure}

\begin{table}[H]
    \centering
    \caption{Distribution of Object Classes in RGB and SEG datasets}
    \label{tab:distribution}
    \begin{tabular}{ccc}
        \toprule
        Class ID & Number of Images (RGB) & Number of Images (SEG) \\
        \midrule
        0  & 20 & 20 \\
        1  & 20 & 20 \\
        2  & 20 & 20 \\
        3  & 20 & 20 \\
        4  & 20 & 20 \\
        5  & 20 & 20 \\
        6  & 20 & 20 \\
        7  & 20 & 20 \\
        8  & 20 & 20 \\
        9  & 20 & 20 \\
        10 & 20 & 20 \\
        11 & 20 & 20 \\
        12 & 20 & 20 \\
        13 & 20 & 20 \\
        14 & 20 & 20 \\
        15 & 20 & 20 \\
        16 & 20 & 20 \\
        17 & 20 & 20 \\
        18 & 20 & 20 \\
        19 & 20 & 20 \\
        \bottomrule
    \end{tabular}
\end{table}

\subsection*{Experimental Conditions}

The primary goal of the experiments is to evaluate the performance of semantic segmentation models for AVs in both centralized and FL frameworks. The experiments aim to compare the accuracy and privacy trade-offs between centralized and FL models, while also assessing the impact of Differential Privacy (DP) on model performance within the FL framework. Additionally, the experiments focus on measuring the communication efficiency and scalability of the FL approach to ensure its applicability in real-world AV environments. The experimentation is divided into two phases:
\begin{itemize}
\item Phase-1 (Centralized Learning): A hybrid UNet-ResNet34 architecture was implemented in a centralized framework to achieve baseline performance in semantic segmentation tasks.
\item Phase-2 (FL with DP): FL was applied across multiple clients (simulating AVs), integrating DP mechanisms to ensure data privacy while maintaining model performance. The training was conducted across 20, 40, and 70 CRs.
\end{itemize}
   
\noindent Both phases were evaluated using identical datasets, model architectures, and hyperparameters to ensure consistent comparison.  Figure 4 illustrates the FL implementation for the RGB and SEG datasets, respectively.
   
Some of the configuration and tools used in the experiments include: high-performance GPUs (NVIDIA Tesla V100) and multiple client machines to simulate decentralized FL environments. TensorFlow, TensorFlow Federated, and TensorFlow Privacy were used to develop the models, while NumPy, Pandas, and Matplotlib handled data processing and visualization.

\begin{figure}[H]
    \centering
    \includegraphics[width=\linewidth]{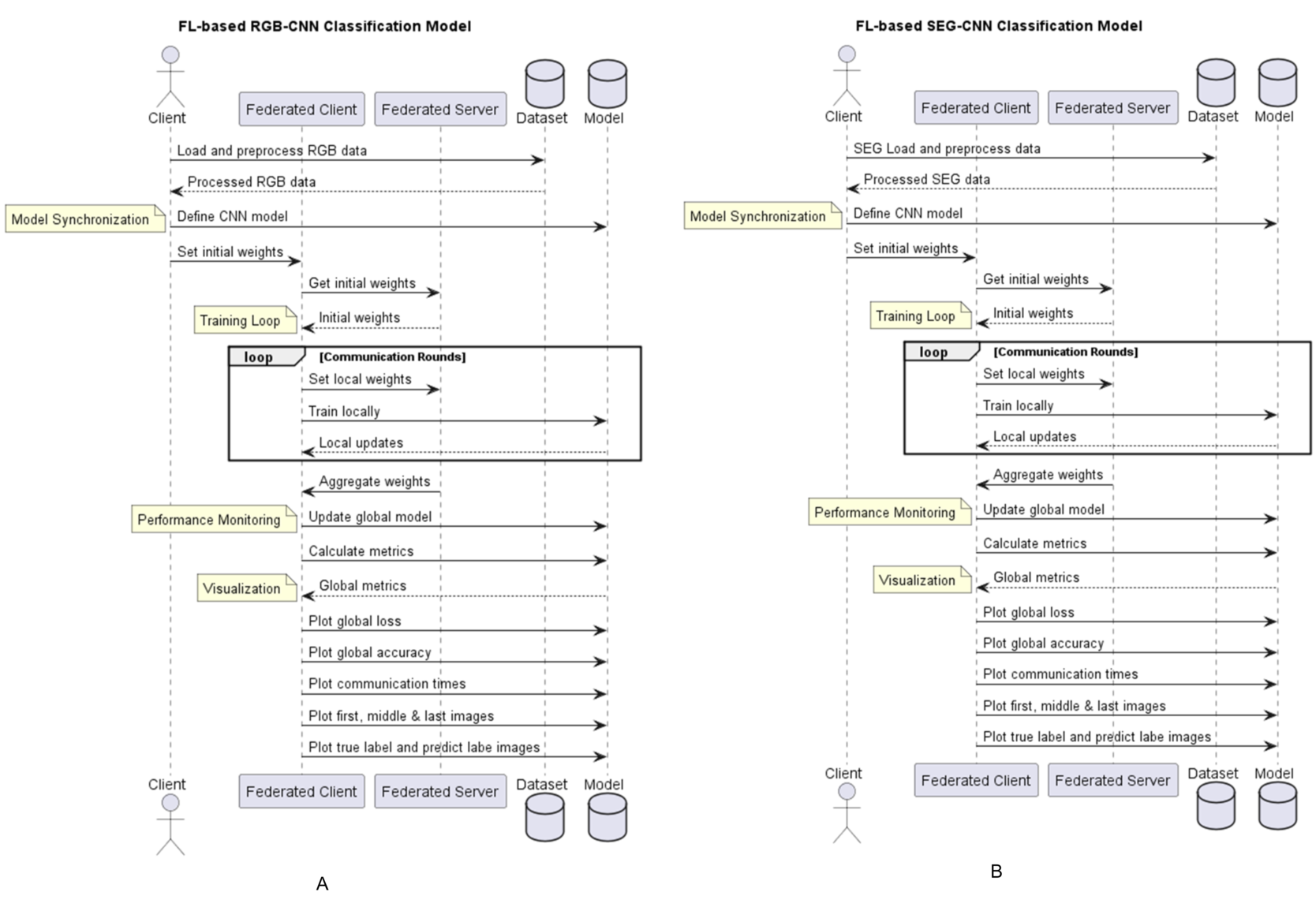}
    \caption{FL-based CNN Classification Model. (A) RGB Dataset, (B) SEG Dataset}
    \label{fig:enter-label}
\end{figure}

\subsection*{FL-CNN Model Parameter Settings on RGB and SEG Datasets}

The experiments were conducted using three configurations based on the number of CRs: 20, 40, and 70. Tables 2 and 3 summarize the model parameters for the RGB dataset, and Tables 4 and 5 for the SEG
283  dataset.
\begin{table}[H]
\centering
\caption{Model Parameters (varied CRs) for FL-CNN for RGB Dataset}
\begin{tabular}{lccc}
\toprule
\textbf{Parameter} & \textbf{CR = 20} & \textbf{CR = 40} & \textbf{CR = 70} \\
\midrule
CRs & 20 & 40 & 70 \\
Samples per Round & 14 & 7 & 4 \\
Batch Size & 32 & 64 & 128 \\
Epochs per Client & 5 & 10 & 20 \\
Displayed Images & First 5 images from dataset & Middle 5 images from dataset & Last 5 images from dataset \\
\bottomrule
\end{tabular}
\end{table}
\begin{table}[H]
\centering
\caption{Model Parameters (Common for all CRs) for FL-CNN for RGB Dataset}
\begin{tabular}{p{6cm}p{9cm}}
\toprule
\textbf{Parameter} & \textbf{Value} \\
\midrule
Mean Pixel Value & [5.06822697, 0., 0.] \\
Standard Deviation Pixel Value & [4.01327451, 0., 0.] \\
Number of Clients & 10 \\
CNN Model Architecture & Sequential \\
& - Conv2D(64, (3, 3), activation='relu', input\_shape=(256, 256, 3)) \\
& - MaxPooling2D((2, 2)) \\
& - Conv2D(32, (3, 3), activation='relu') \\
& - MaxPooling2D((2, 2)) \\
& - Conv2D(32, (3, 3), activation='relu') \\
& - MaxPooling2D((2, 2)) \\
& - Flatten() \\
& - Dense(64, activation='relu') \\
& - Dropout(0.5) \\
& - Dense(num\_classes, activation='softmax') \\
Optimizer & Adam (learning\_rate=0.0001) \\
Client Training Optimizer & RMSProp \\
DP Epsilon & 0.1 \\
Loss Function & sparse\_categorical\_crossentropy \\
Metrics & Accuracy \\
Train-Test Split & 70\% train, 30\% test \\
Image Resize Dimensions & (256, 256) \\
\bottomrule
\end{tabular}
\end{table}
\begin{table}[H]
\centering
\caption{Model Parameters (varied CRs) for FL-CNN for SEG Dataset}
\begin{tabular}{p{5cm}p{3cm}p{3cm}p{3cm}}
\toprule
\textbf{Parameter} & \textbf{CR = 20} & \textbf{CR = 40} & \textbf{CR = 70} \\
\midrule
CRs & 20 & 40 & 70 \\
Samples per Round & 14 & 7 & 4 \\
Batch Size & 32 & 64 & 128 \\
Epochs per Client & 5 & 10 & 20 \\
Displayed Images & First 5 images from dataset & Middle 5 images from dataset & Last 5 images from dataset \\
\bottomrule
\end{tabular}
\end{table}
\begin{table}[H]
\centering
\caption{Model Parameters (Common for all CRs) for FL-CNN for SEG Dataset}
\begin{tabular}{p{7cm}p{8cm}}
\toprule
\textbf{Parameter} & \textbf{Value} \\
\midrule
Mean Pixel Value & [5.06822697, 0., 0.] \\
Standard Deviation Pixel Value & [4.01327451, 0., 0.] \\
Number of Clients & 10 \\
CNN Model Architecture & Sequential \\
& - Conv2D(64, (3, 3), activation='relu', input\_shape=(256, 256, 3)) \\
& - MaxPooling2D((2, 2)) \\
& - Conv2D(32, (3, 3), activation='relu') \\
& - MaxPooling2D((2, 2)) \\
& - Conv2D(32, (3, 3), activation='relu') \\
& - MaxPooling2D((2, 2)) \\
& - Flatten() \\
& - Dense(64, activation='relu') \\
& - Dropout(0.5) \\
& - Dense(num\_classes, activation='softmax') \\
Optimizer & Adam (learning\_rate=0.0001) \\
Client Training Optimizer & RMSprop \\
DP Epsilon & 0.1 \\
Loss Function & sparse\_categorical\_crossentropy \\
Metrics & Accuracy \\
Train-Test Split & 70\% train, 30\% test \\
Image Resize Dimensions & (256, 256) \\
\bottomrule
\end{tabular}
\end{table}

\subsection*{Performance Evaluation Metrics}
The performance of the FL models with DP was evaluated using the following metrics:

\begin{enumerate}
\item Accuracy: Measures the proportion of correctly classified pixels (Equation 7).
    
\item Mean IoU: Represents the average overlap between predicted and ground truth segments across classes (Equation 8).
    
\item Specificity and Sensitivity (Recall): Evaluate the model’s ability to correctly classify positive and negative segments (Equations 9 and 10).

\item Precision and F1 Score: Measure the balance between precision and recall (Equations 11 and 12).
    
\item Communication Efficiency: Assesses the time and resources required to transmit model updates between clients and the server.
\end{enumerate}
\begin{equation}
    \text{Accuracy} = \frac{\text{Number of Correctly Classified Pixels}}{\text{Total Number of Pixels}}
\end{equation}
\begin{equation}
    \text{Mean IoU} = \frac{1}{C} \sum_{c=1}^C \text{IoU}_c
\end{equation}
\begin{equation}
    \text{Sensitivity (Recall)} = \frac{\text{True Positive}}{\text{True Positive} + \text{False Negative}}
\end{equation}
\begin{equation}
    \tilde{w}_i^{t+1} = w_i^{t+1} + \mathcal{N}(0, \sigma^2)
\end{equation}
\begin{equation}
    \text{Precision} = \frac{\text{True Positive}}{\text{True Positive} + \text{False Positive}}
\end{equation}
\begin{equation}
    \text{F1 Score} = 2 \times \frac{\text{Precision} \times \text{Recall}}{\text{Precision} + \text{Recall}}
\end{equation}

\subsection*{Effect of Privacy Budget on Performance}

The impact of varying the privacy budget ($\epsilon$) on model performance was evaluated by testing with different values of $\epsilon = 0.1$, $\epsilon = 0.5$, and $\epsilon = 1.0$. As the privacy budget increases, the noise added to the model updates decreases, leading to better accuracy but reduced privacy protection. Table 6 summarizes the impact on segmentation accuracy and Mean IoU. This table highlights the trade-off between privacy and performance. While a smaller $\epsilon$ provides stronger privacy, it slightly reduces model accuracy, as seen with $\epsilon = 0.1$, where the accuracy is 86.5\%. On the other hand, larger values of $\epsilon$ lead to higher accuracy but offer weaker privacy guarantees.
\begin{table}[H]
\centering
\caption{Effect of Privacy Budget on Accuracy and Mean IoU}
\begin{tabular}{ccc}
\toprule
\textbf{Privacy Budget ($\varepsilon$)} & \textbf{Accuracy (\%)} & \textbf{Mean IoU (\%)} \\
\midrule
0.1 & 86.5 & 78.0 \\
0.5 & 88.0 & 80.5 \\
1.0 & 89.2 & 82.1 \\
\bottomrule
\end{tabular}
\end{table}

\subsection*{Comparison with State-of-the-Art Methods}

In this subsection, we compare our proposed FL framework with DP to existing state-of-the-art methods for semantic segmentation in AVs. This comparison emphasizes key aspects such as handling non-IID data, communication efficiency, and data privacy.
Table 7 outlines the advantages of our method over centralized DL models and traditional FL approaches, which lack DP integration.

As shown in Table 7, centralized DL models such as UNet and ResNet have high computational power but perform poorly in terms of handling non-IID data and communication efficiency. These models require centralized data aggregation, which raises privacy concerns and limits scalability for real-time AV environments. In contrast, our FL-based method with DP offers a decentralized solution, effectively addressing non-IID data across multiple vehicles while maintaining strong data privacy through the DP mechanism.

Additionally, our proposed method ensures communication efficiency by reducing the frequency and volume of data transmitted between clients and the central server. Traditional FL models without DP struggle to achieve the same level of privacy protection, and while they provide moderate communication efficiency, they do not fully address the privacy risks associated with centralized data aggregation.

By integrating DP into FL, our approach not only safeguards sensitive data but also maintains high accuracy and segmentation performance in real-world AV applications, ensuring both privacy and performance are balanced effectively.

\begin{table}[H]
\centering
\caption{Comparison of Proposed Method with State-of-the-Art Methods}
\begin{tabular}{p{5cm}p{3cm}p{3.5cm}p{3.5cm}}
\toprule
\textbf{Method} & \textbf{Non-IID Handling} & \textbf{Communication Efficiency} & \textbf{Data Privacy (DP)} \\
\midrule
Centralized DL Models (e.g., UNet, ResNet) & No & Low & None \\
Standard FL (No DP) & Yes (limited) & Moderate & None \\
FL with DP (Proposed Method) & Yes (robust) & High & Strong (DP mechanism) \\
PVDM-YOLOv8I\cite{Tahir2024} & No & Low & None \\
\bottomrule
\end{tabular}
\end{table}

\subsection*{Phenomena, Causes, and Recommendations}

Several phenomena were observed during the experiments. Increasing the number of communication rounds (CRs) led to improved accuracy, but at the cost of added latency. This was primarily due to the larger volume of data exchanged between clients and servers in higher CRs. To address this, it is recommended to optimize communication protocols to reduce latency without compromising accuracy. Additionally, the use of Differential Privacy (DP) resulted in a slight reduction in accuracy, which can be attributed to the noise introduced by the DP mechanisms. To mitigate this trade-off, further optimization of DP parameters is suggested.

\section*{Results and Discussion}

\subsection*{Phase-1 Results}

\subsubsection*{Performance Analysis of the Hybrid UNet-ResNet34 Model}

In Phase-1, we implemented a hybrid UNet-ResNet34 model for semantic segmentation on the CARLA simulator dataset. The training process involved extensive data pre-processing and augmentation to ensure the model was exposed to diverse scenarios, enhancing its robustness and generalization capabilities.
\begin{enumerate}
\item  Accuracy: The hybrid model achieved an accuracy of 92.5\% on the validation set, indicating its high capability to correctly classify the majority of the pixels in the segmentation tasks.
 
\item Mean IoU: The model attained a Mean IoU of 0.78, demonstrating its efficiency in overlapping the predicted segments with the ground truth labels.
  
\item Sensitivity (Recall): The sensitivity of the model was calculated at 0.85, reflecting its effectiveness in identifying true positive segments.
 
\item Specificity: The specificity stood at 0.91, showing the model’s proficiency in correctly identifying negative segments.
 
\item Precision: The precision value of the model was 0.87, highlighting its accuracy in predicting positive segments.

\end{enumerate}
The graph shown in Figure 5 presents a comparative analysis of the Mean IoU performance of four different models: FCN, UNet, ENet, and the Phase-1 Framework. The Mean IoU is a standard metric used to evaluate the accuracy of semantic segmentation models, where higher values indicate better performance. The model proposed in Phase-1 consistently outperforms the other models across all categories. For unlabelled regions, the Phase-1 model achieves a Mean IoU of 0.35, compared to 0.3 for ENet, 0.25 for UNet, and 0.1 for FCN. In the building category, the proposed model scores 0.65, surpassing ENet’s 0.6, UNet’s 0.55, and FCN’s 0.45. For fences, the proposed model reaches 0.5, higher than ENet’s 0.45, UNet’s 0.4, and FCN’s 0.25. Similarly, for the ’Other’ category, it scores 0.3, compared to ENet’s 0.25, UNet’s 0.2, and FCN’s 0.1. The Phase-1 model also excels in pedestrian segmentation with a Mean IoU of 0.35, poles at 0.3, road lines at 0.75, roads at 0.85, sidewalks at 0.6, vegetation at 0.75, cars at 0.7, walls at 0.55, and traffic signs at 0.4, consistently outperforming the other models in each category. This indicates that the enhancements in the Phase-1 model significantly improve its performance in semantic segmentation tasks, making it more accurate and reliable for AV applications.
 
Key Findings of Phase-1: The proposed Phase-1 model’s accuracy was 0.92, compared to 0.89 for UNet, 0.85 for FCN, and 0.88 for ENet. The Mean IoU improved significantly to 0.75 from 0.68 (UNet), 0.65 (FCN), and 0.70 (ENet). The recall score was 0.85, indicating the model’s capability in detecting relevant positive instances. The specificity stood at 0.91, and the precision value of the model was 0.87.

\begin{figure}[H]
\centering
\includegraphics[width=0.6\linewidth]{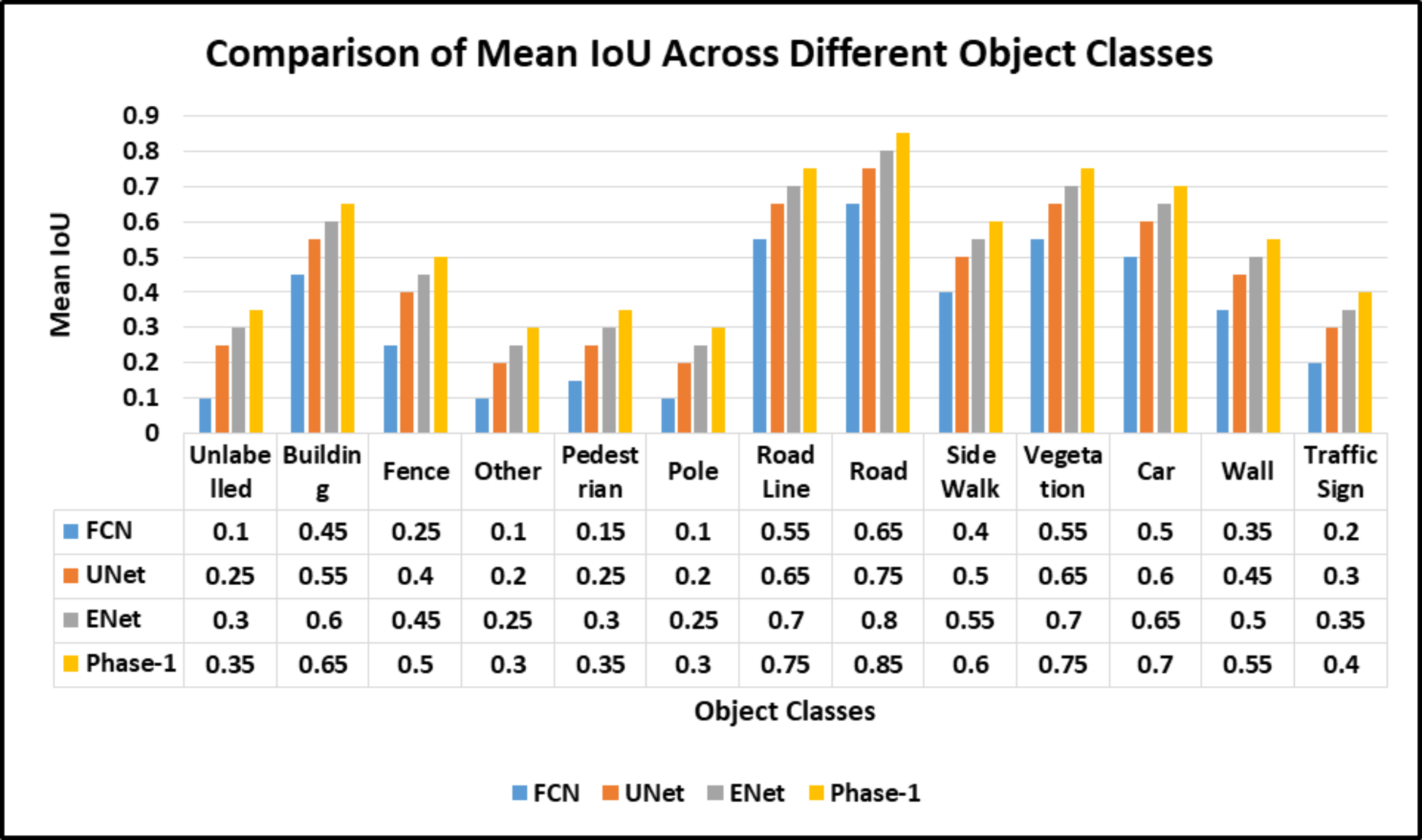}
\caption{Comparison of Mean IoU Across Different Object Classes}
\label{fig:enter-label}
\end{figure}

\subsubsection*{Visual Results Demonstrating the Segmentation Capability}

The segmentation results of the hybrid UNet-ResNet34 model, shown in figure 6, closely match the ground truth annotations. The figure illustrates the original input image, the true segmentation mask, and the model’s segmented output for object classes like buildings, poles, roadlines, roads, sidewalks, and cars, with color-coded segments. This visual demonstrates the model’s effectiveness in accurately segmenting diverse elements within the driving environment, highlighting its utility for autonomous driving applications.
\begin{figure}[H]
\centering
\subfigure{
\includegraphics[width=0.7\textwidth]{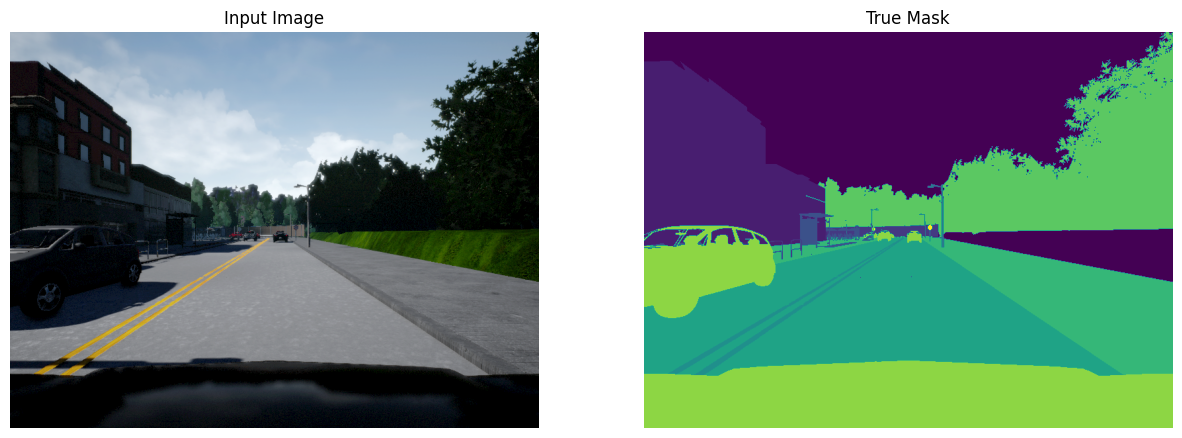}}
\subfigure{\includegraphics[width=0.35\textwidth]{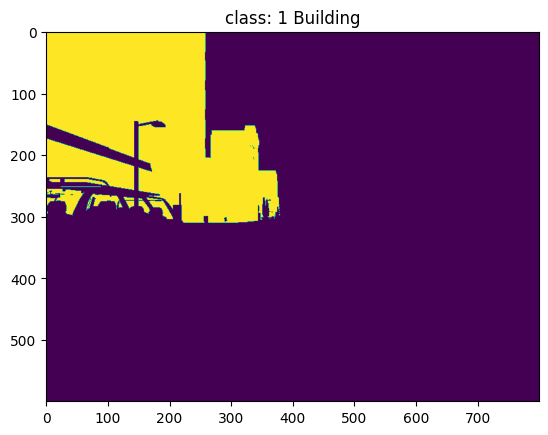}
\includegraphics[width=0.35\textwidth]{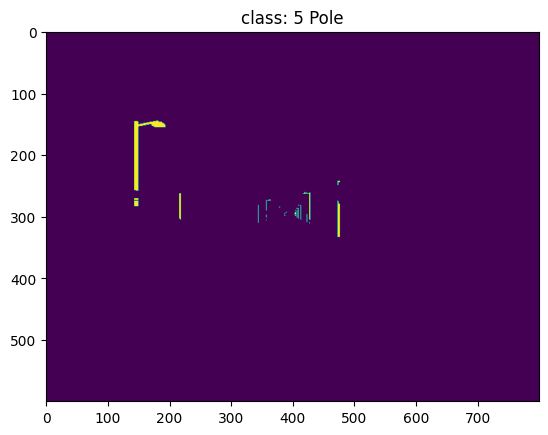}}
\subfigure{\includegraphics[width=0.35\textwidth]{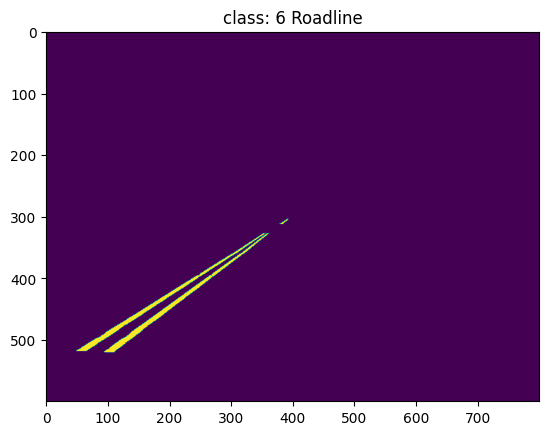}
\includegraphics[width=0.35\textwidth]{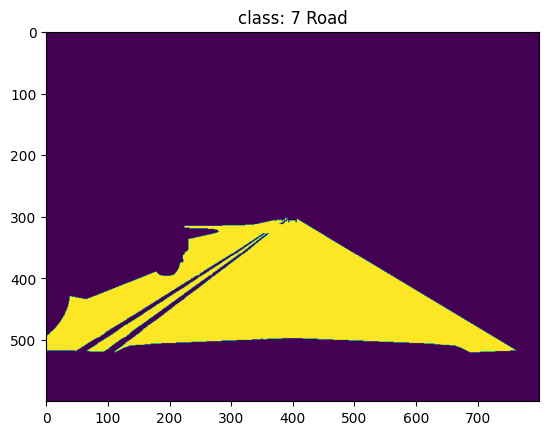}
    }
\subfigure{
\includegraphics[width=0.35\textwidth]{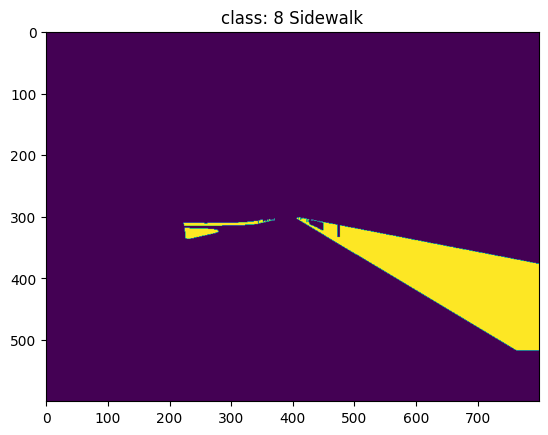}
\includegraphics[width=0.35\textwidth]{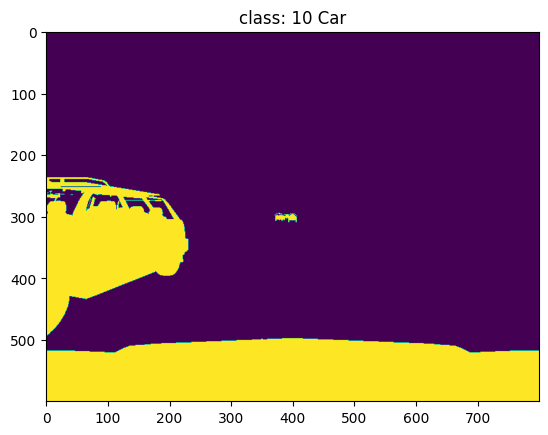}
     }
\caption{Segmentation results obtained from Phase-1 experimentation using the Hybrid UNet-ResNet34 Model}
    \label{fig:merged}
\end{figure}

\subsubsection*{Comparative Analysis of Models used in Phase-1}

Figure 7 compares the performance of various semantic segmentation models, including UNet, FCN, ENet, and the proposed Phase-1 model. The metrics evaluated include Accuracy, Mean IoU, Sensitivity, Specificity, Precision, and Recall. The proposed model consistently outperforms others, particularly in Accuracy (0.93) and Recall (0.90). Mean IoU shows the most significant variation, where the proposed model demonstrates a clear advantage. This analysis confirms the superior segmentation accuracy and robustness of the hybrid UNet-ResNet34 model.

\begin{figure}[H]
    \centering
    \includegraphics[width=0.6\linewidth]{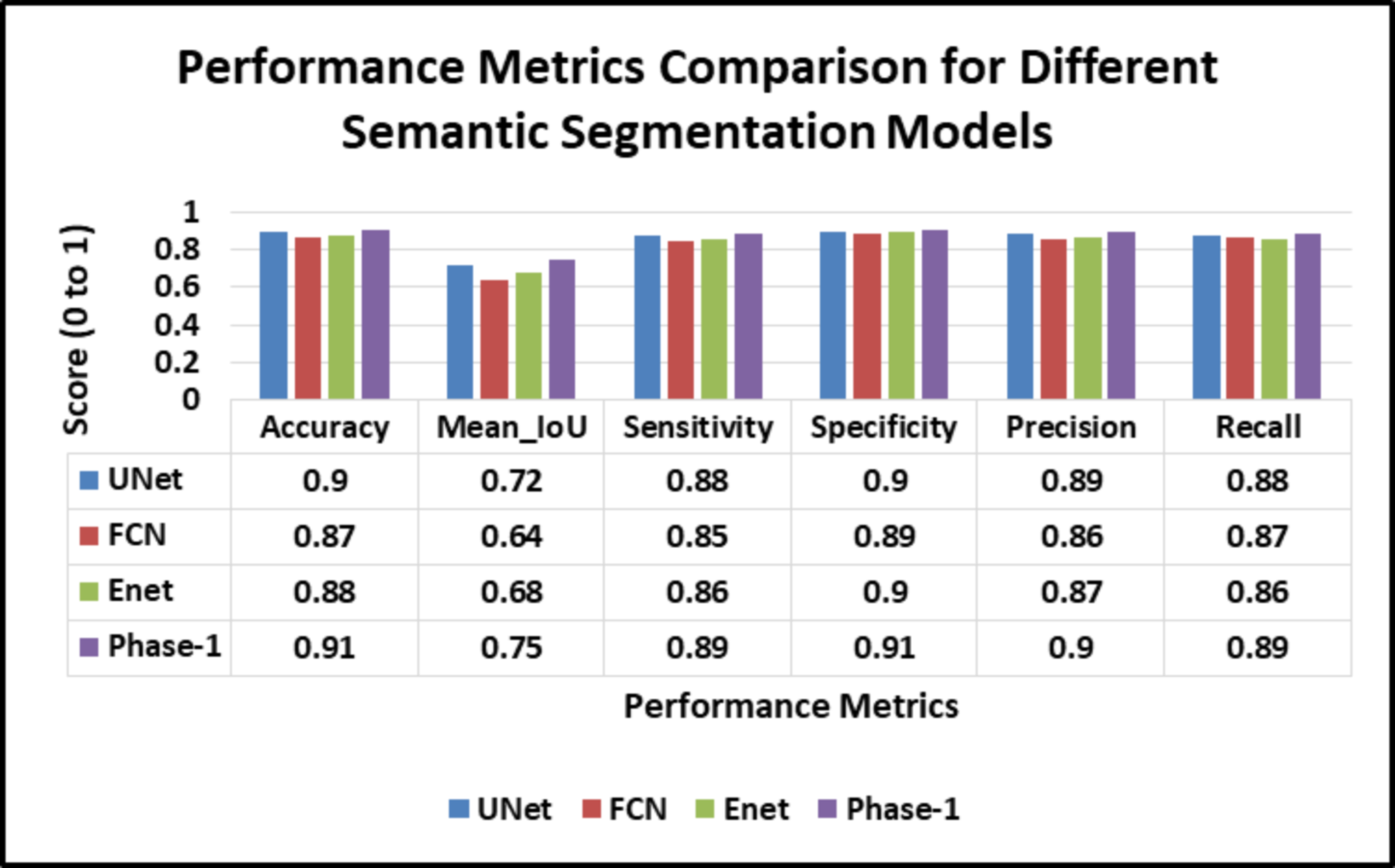}
    \caption{Performance Metrics Comparison for Different Semantic Segmentation Models used in Phase-1}
    \label{fig:enter-label}
\end{figure}

\subsection*{Phase-2 Results (Performance Analysis of FL Models)}

In Phase-2, FL with secure deep CNNs was implemented to improve segmentation performance while addressing privacy concerns. The models were evaluated over 20, 40, and 70 CRs on RGB and SEG datasets.

\subsubsection*{FL-based CNN Performance on RGB Dataset}

Figure 8 illustrates the performance of the FL-based CNN across different CRs. The results show a consistent reduction in global loss, reaching minimal values by the 70th round. Correspondingly, global accuracy improves significantly, achieving nearly 98\% after 70 CRs, indicating robust learning. Communication time remains within a manageable range across all CRs, demonstrating stable and scalable performance even with extended rounds.

\begin{figure}[H]
    \centering
    \includegraphics[width=1\linewidth]{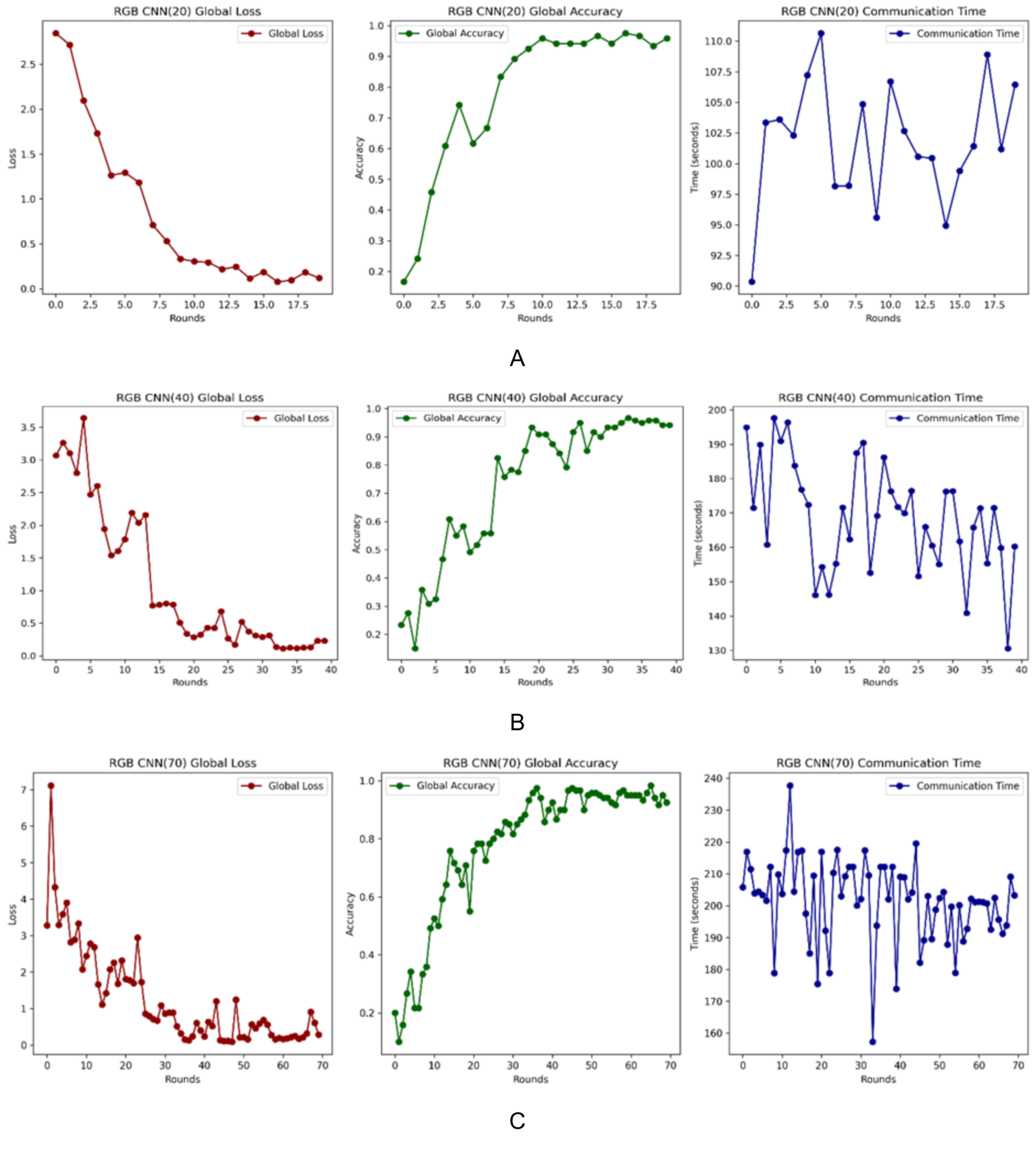}
    \caption{Global Loss, Global Accuracy, and Communication Time of FL-based CNN for RGB Dataset (A) 20 CRs, (B) 40 CRs, (C) 70 CRs}
    \label{fig:enter-label}
\end{figure}

\subsubsection*{Analysis of Performance Metrics across Different CRs on RGB Dataset}

The global loss consistently decreased across 20, 40, and 70 CRs, starting at 2.8, 3.5, and 7.0, respectively, and dropping to 0.0 in all cases, indicating effective learning. Global accuracy showed a steady improvement, with 20 CRs rising from 0.2 to 0.98, 40 CRs following a similar trend, and 70 CRs achieving smoother gains, all stabilizing at 0.98. Communication time, however, increased with more CRs, starting at 90 seconds for 20 CRs, 140 seconds for 40 CRs, and 240 seconds for 70 CRs, highlighting a trade-off between prolonged training and communication overhead. Overall, the results demonstrate significant reductions in loss and gains in accuracy with extended training rounds, while also indicating the need to optimize communication time in FL environments.

\subsubsection*{Visual Results Demonstrating the Segmentation Capability on RGB Dataset}

The sub-figure A in figure 9 shows the segmentation results for RGB CNN with 20 CRs, accurately identifying roads, buildings, and vegetation. The comparison in sub-figure A of figure 9 between true and predicted labels highlights the model’s precision and recall. The sub-figures B of figures 9 display improved segmentation for 40 CRs, where more refined outputs are observed due to enhanced training, particularly in complex scenes. For 70 CRs, as shown in sub-figures C of figures 9, segmentation accuracy further improves, even in challenging conditions, with near-perfect alignment between true and predicted labels, showcasing the model’s generalization capabilities. The results in figures 5 show clear improvements in segmentation and efficiency as training rounds increase, underscoring the benefits of extended FL training for better performance.

\begin{figure}[H]
    \centering
    \includegraphics[scale=5]{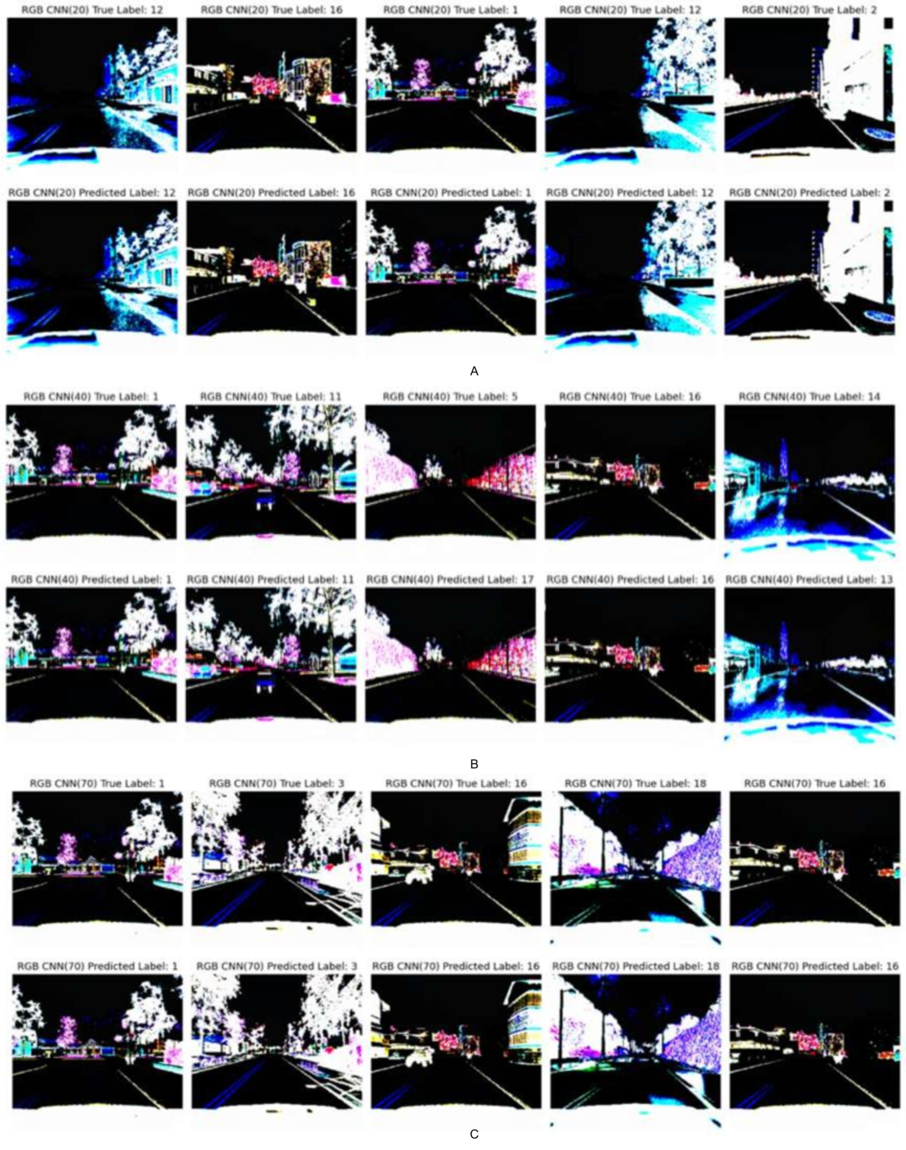}
    \caption{True vs. Predicted Labels of FL-CNN for RGB Dataset. (A) 20 CRs, (B) 40 CRs, (C) 70 CRs}
    \label{fig:enter-label}
\end{figure}

\subsubsection*{Analysis of Segmentation Outputs and True vs. Predicted Labels across Different CRs on RGB Dataset}

The segmentation outputs at 20, 40, and 70 CRs show a clear progression in model performance. At 20 CRs (Images 1-5), the segmentation is basic, with some inaccuracies in delineating elements like road lines and vehicles, indicating the need for more training. At 40 CRs (Images 199-203), accuracy improves, with clearer distinctions and more precise segmentation masks. The model captures variations better, leading to more robust feature learning. At 70 CRs (Images 396-400), the outputs are highly refined, with precise identification of elements, demonstrating the benefits of extended training in enhancing model precision and accuracy.

The comparison between true and predicted labels shows a consistent learning trend. At 20 CRs, there is some alignment but with noticeable discrepancies, indicating the need for more training. At 40 CRs, the alignment improves significantly, reducing misclassification and showing the model’s enhanced ability to generalize. At 70 CRs, the labels align closely, with minimal errors, demonstrating high accuracy and reliability due to extensive training. This highlights the advantage of extended CRs in FL, enabling robust performance and high accuracy. The analysis shows a progressive improvement with increasing CRs, from foundational learning at 20 CRs to refined performance at 70 CRs. This emphasizes the importance of adequate CRs in achieving precise and reliable segmentation through iterative training in FL models.

\subsubsection*{FL-based CNN Performance on SEG Dataset}

The graph in sub-figure A of Figure 10 shows a noticeable decrease in global loss over 20 CRs for the SEG dataset. The loss starts high and decreases steadily, reflecting effective learning with each round. The second graph shows global accuracy rising to 85\% by the 20th round, indicating the model’s ability to learn and generalize from the data. The communication time in the third graph fluctuates but stays within an efficient range, showing a stable FL process. As shown in sub-figure B of Figure 10, the global loss decreases more significantly over 40 CRs, with a steeper decline compared to 20 CRs. The global accuracy reaches 90\%, demonstrating improved learning with extended CRs. The communication time remains manageable despite more fluctuations. The graph in sub-figure C of Figure 10 shows a sharp reduction in global loss over 70 CRs, indicating strong convergence. The second graph shows accuracy reaching 95\%, nearly optimal performance due to the extended training. Communication time varies more but stays within a reasonable range, indicating that the FL approach remains efficient even with extended CRs.

\begin{figure}[H]
    \centering
    \includegraphics[width=0.7\linewidth]{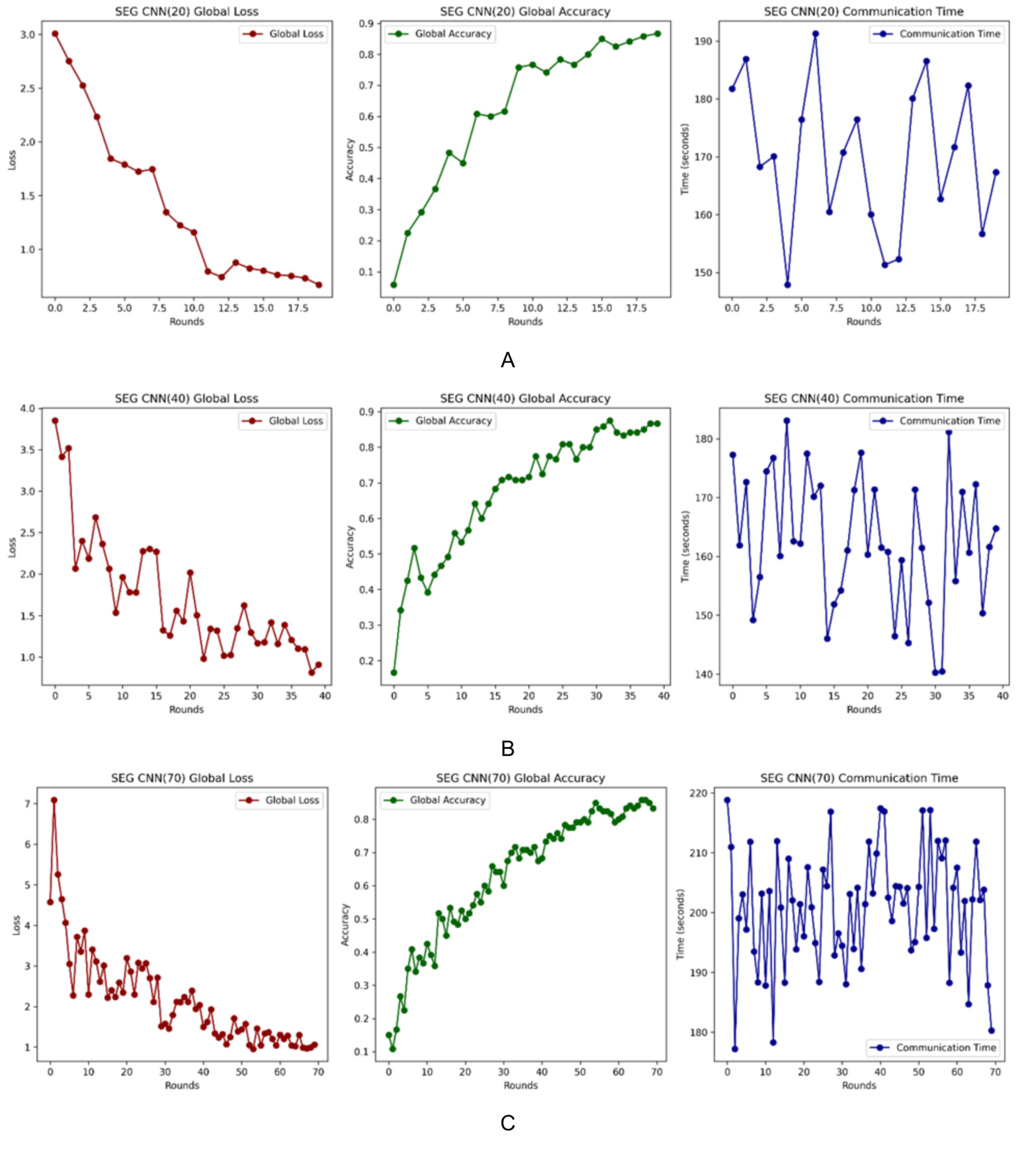}
    \caption{Global Loss, Global Accuracy, and Communication Time of FL-based CNN for SEG Dataset. (A) 20 CRs, (B) 40 CRs, (C) 70 CRs}
    \label{fig:enter-label}
\end{figure}

\subsubsection*{Analysis of Performance Metrics across Different CRs on SEG Dataset}

The global loss consistently declines across 20, 40, and 70 CRs, reflecting effective model learning and convergence, with higher initial losses dropping near zero as training progresses. For 20 CRs, the global loss begins high and decreases to near zero, indicating stable training; at 40 CRs, the loss starts higher but drops significantly, while the 70 CRs model, starting with the highest loss, achieves a marked reduction, confirming the benefit of extended rounds. Global accuracy shows clear improvement across all CRs, with 20 CRs rising rapidly from a low starting point, 40 CRs benefiting from incremental learning with some fluctuations, and 70 CRs stabilizing at a higher level, confirming the advantage of extended training rounds. Communication time varies, with 20 CRs showing more pronounced fluctuations, 40 CRs displaying a consistent upward trend, and 70 CRs experiencing a gradual rise, highlighting the trade-off between increased accuracy and communication overhead in FL environments.

\subsubsection*{Visual Results Demonstrating the Segmentation Capability on SEG Datasets}

The sub-figure A in figure 11 presents segmentation results for SEG CNN with 20 CRs, showing accurate segmentation of objects like buildings, fences, and vegetation, demonstrating the model’s capacity for diverse environments. The sub-figure B in figure 12 compares true and predicted labels, showing close alignment, reflecting high precision and recall. The sub-figures B in figures 11 and 12 show segmentation outputs and label comparisons for 40 CRs. Enhanced training improves segmentation, especially in complex scenes, with predicted labels closely matching the ground truth. The sub-figures C of figures 11 and 12 illustrate results for 70 CRs, showing highly accurate segmentation even in challenging conditions. Label comparison confirms near-perfect alignment, indicating strong generalization from extensive training.

\begin{figure}[H]
    \centering
    \includegraphics[width=0.7\linewidth]{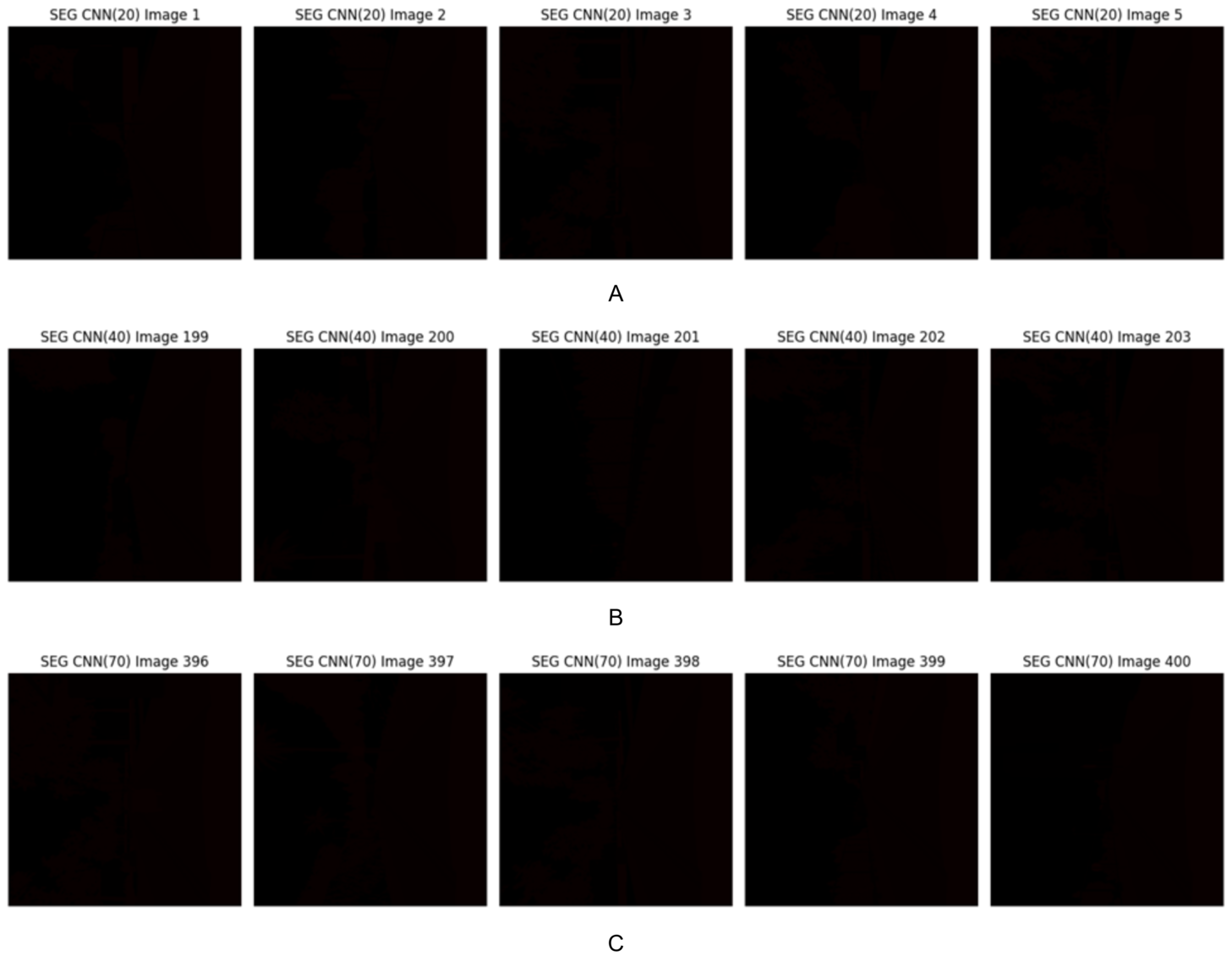}
    \caption{Segmentation Outputs of FL-CNN for SEG Dataset. (A) 20 CRs, (B) 40 CRs, (C) 70 CRs}
    \label{fig:enter-label}
\end{figure}

\begin{figure}[H]
    \centering
    \includegraphics[width=0.7\linewidth]{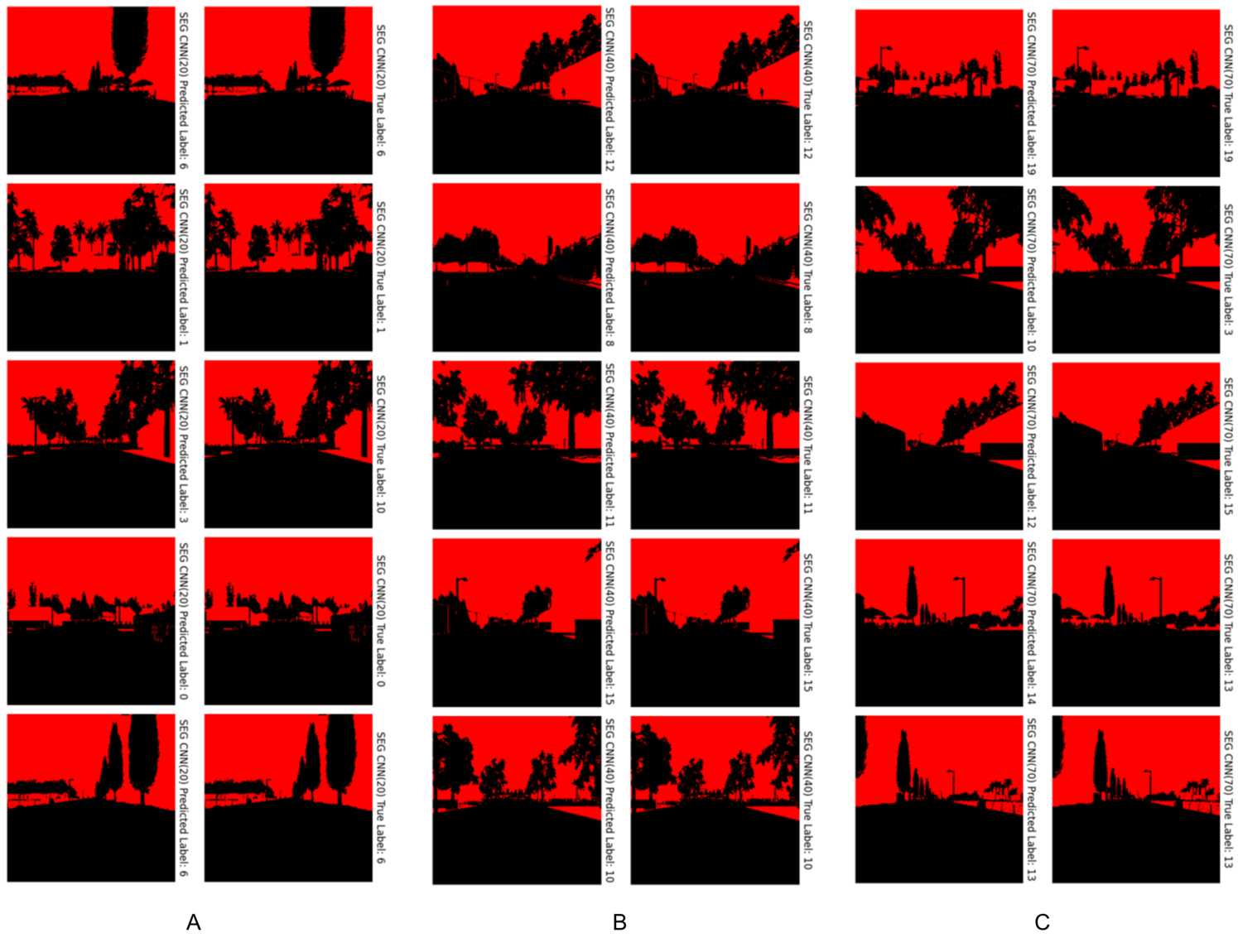}
    \caption{True vs. Predicted Labels of FL-CNN for SEG Dataset. (A) 20 CRs, (B) 40 CRs, (C) 70 CRs}
    \label{fig:enter-label}
\end{figure}

\subsubsection*{Analysis of Segmentation Outputs and True vs. Predicted Labels across Different CRs on SEG Dataset}

For 20 CRs, the segmentation outputs show basic accuracy but some elements are not clearly defined, indicating the need for more training. With 40 CRs, segmentation accuracy and clarity improve, leading to more precise outputs. At 70 CRs, segmentation is highly refined, showing clear delineation of various elements, reflecting the benefits of extended rounds.

For 20 CRs, there are still discrepancies between true and predicted labels, with some misclassifications. At 40 CRs, the alignment improves significantly, with reduced misclassification. For 70 CRs, the alignment is nearly perfect, with minimal misclassification, showing the model’s strong performance with extended training. The analysis highlights the progressive improvement in segmentation outputs and label alignment with extended CRs, underscoring the importance of sufficient rounds for optimal model performance. 

\subsection*{Visual Results Demonstrating the Segmentation Capability of FL-based CNN Models on RGB and SEG Datasets}

The sub-figure A of figure 13 shows the entropy distribution for RGB images, peaking around 6.5, indicating moderate complexity. The sub-figure B of figure 13 shows entropy for the SEG dataset, with values mostly between 1.2 and 1.5. The sub-figure A of figure 14 shows the RGB dataset has mean values around 85-90 across channels. Whereas, in the sub-figure B of figure 14, the SEG dataset shows lower mean values, reflecting its grayscale nature. In sub-figure A of figure 15, the RGB dataset shows high contrast values. The sub-figure B of figure 15 shows the SEG dataset with high homogeneity and correlation, indicating uniformity. The sub-figure A of figure 16 shows the RGB dataset with peaks at low and high intensities. Whereas, the sub-figure B of figure 16 shows the SEG dataset with mostly low intensities, indicating darker images. The sub-figures A and B of figure 17 show RGB and SEG dataset augmentations, including rotations and lighting changes, enhancing model robustness. The sub-figures A and B of figure 18 show edge detection results for RGB and SEG datasets, highlighting critical boundaries for segmentation.

\begin{figure}[H]
    \centering
    \includegraphics[width=0.7\linewidth]{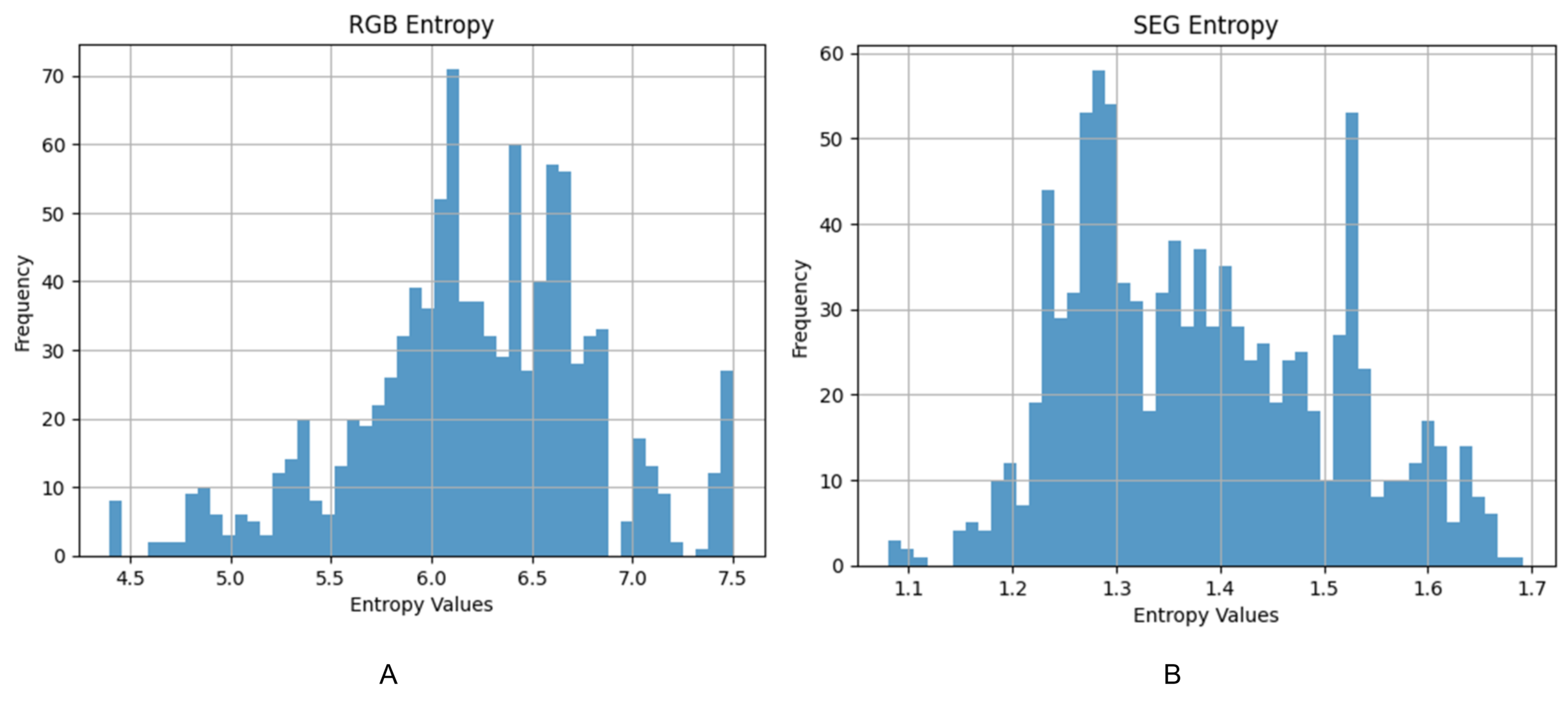}
    \caption{Entropy Distribution. (A) RGB Dataset, (B) SEG Dataset}
    \label{fig:enter-label}
\end{figure}

\begin{figure}[H]
    \centering
    \includegraphics[width=0.9\linewidth]{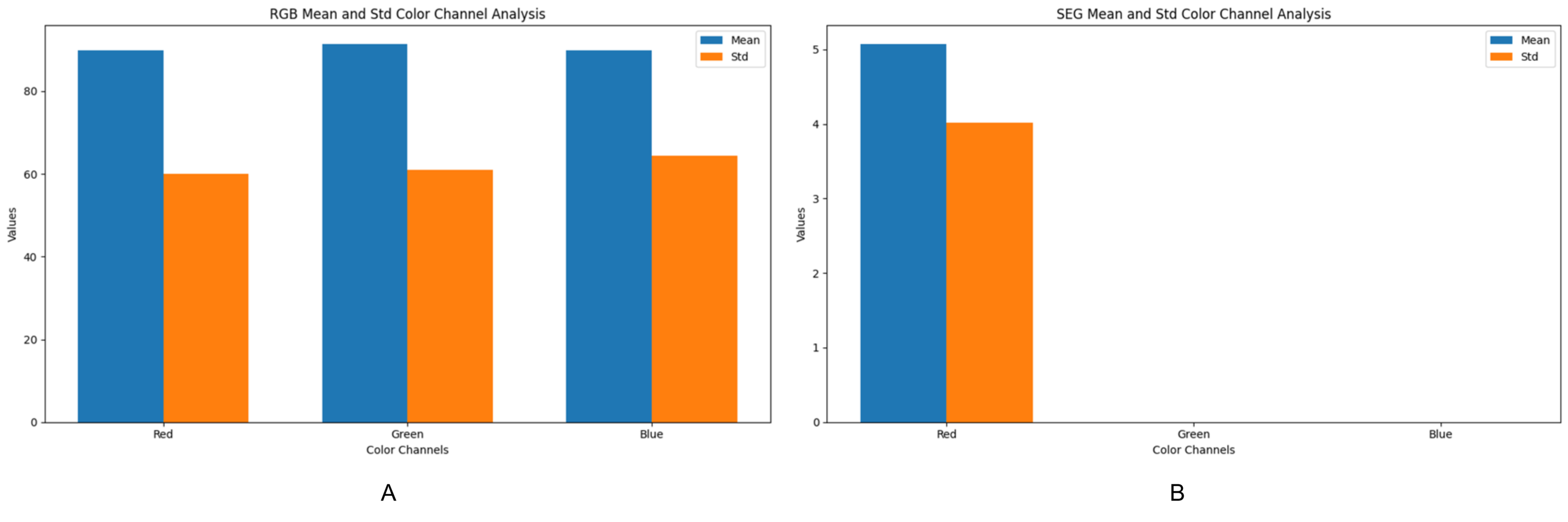}
    \caption{Mean and Standard Deviation Color Channel Analysis. (A) RGB Dataset, (B) SEG Dataset}
    \label{fig:enter-label}
\end{figure}

\begin{figure}[H]
    \centering
    \includegraphics[width=1.1\linewidth]{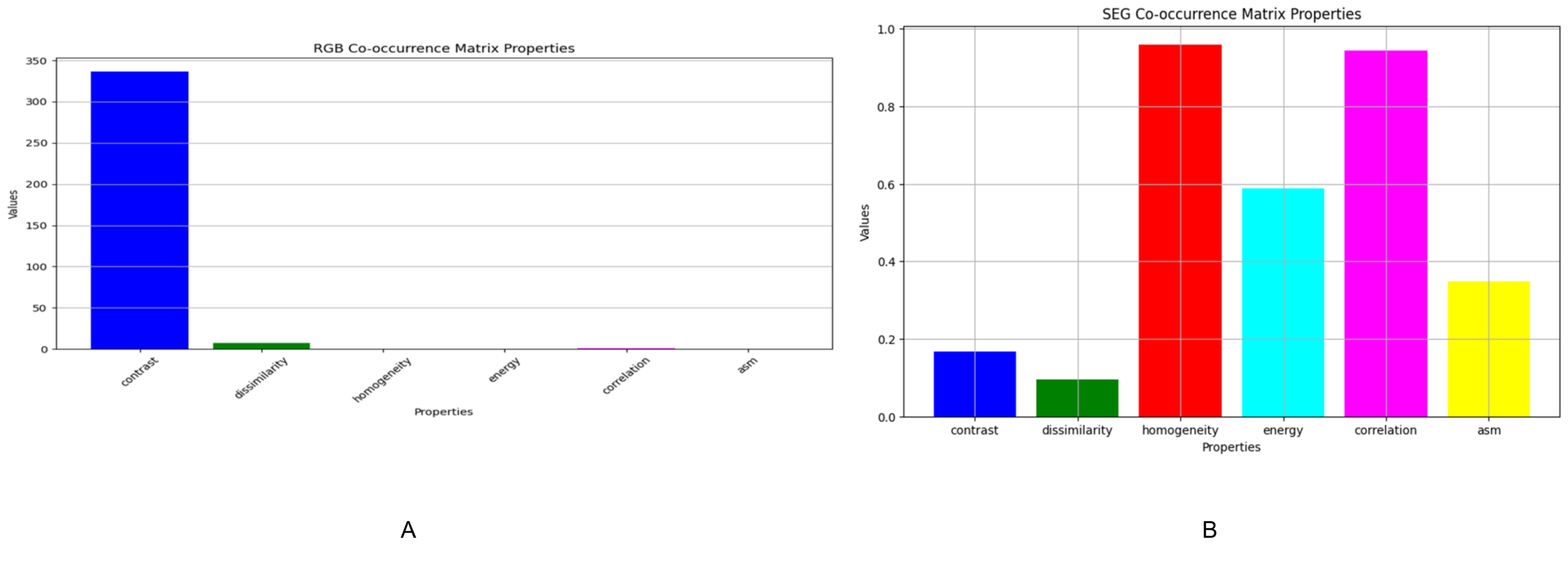}
    \caption{Co-occurrence Matrix Properties. (A) RGB Dataset, (B) SEG Dataset}
    \label{fig:enter-label}
\end{figure}

\begin{figure}[H]
    \centering
    \includegraphics[width=1.1\linewidth]{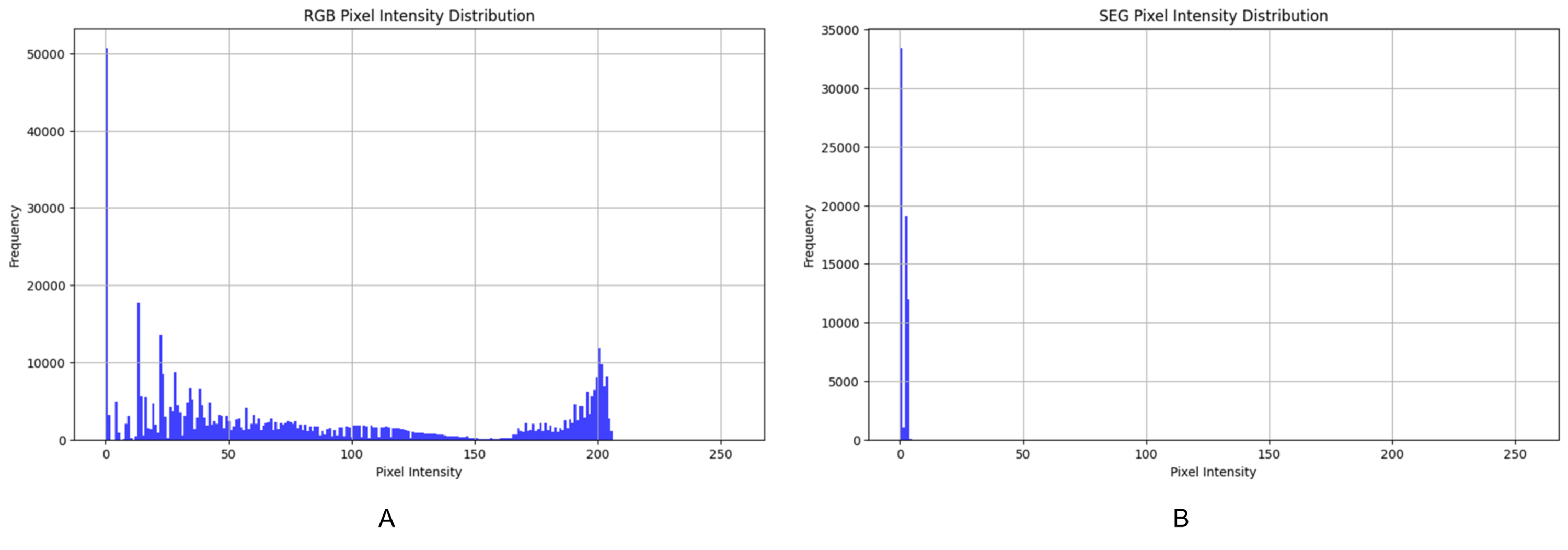}
    \caption{Pixel Intensity Distribution. (A) RGB Dataset, (B) SEG Dataset}
    \label{fig:enter-label}
\end{figure}

\begin{figure}[H]
    \centering
    \includegraphics[width=0.5\linewidth]{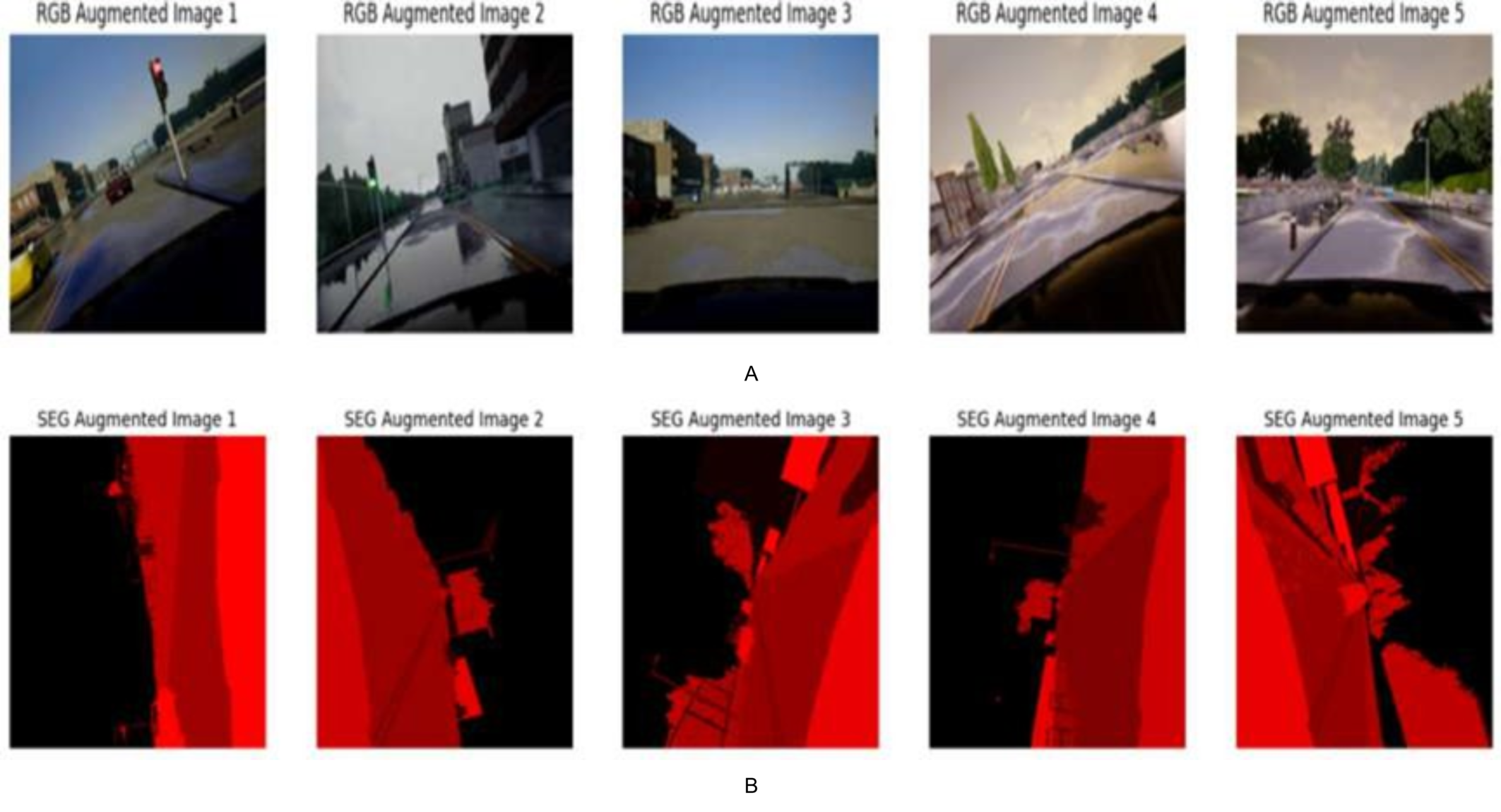}
    \caption{Augmented Images. (A) RGB Dataset, (B) SEG Dataset}
    \label{fig:enter-label}
\end{figure}

\begin{figure}[H]
    \centering
    \includegraphics[width=0.7\linewidth]{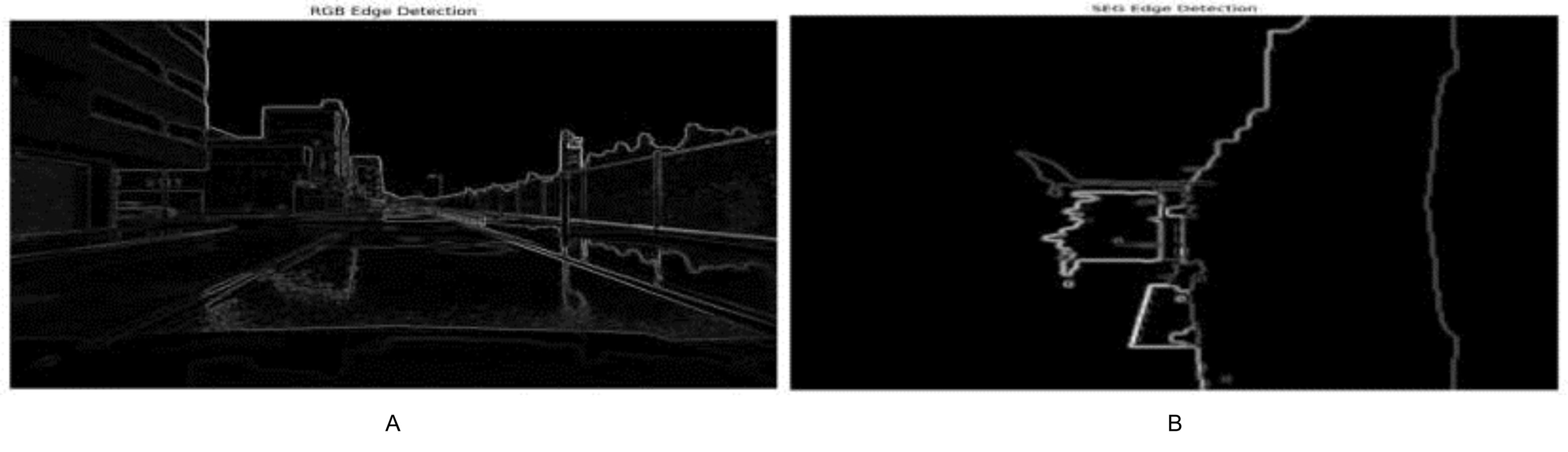}
    \caption{Edge Detection. (A) RGB Dataset, (B) SEG Dataset}
    \label{fig:enter-label}
\end{figure}
\subsubsection*{Impact of DP on FL-based CNN Model Performance on RGB and SEG Datasets Global Loss, Global Accuracy, and Communication Time}

Figure 19 shows the impact of DP across CRs. For RGB, loss decreases from 3.1 to 2.6, while accuracy increases from 0.05 to 0.35. SEG results show loss reducing from 3.0 to 2.86, and accuracy increasing from 0.04 to 0.18. Communication time varies but remains within reasonable ranges, reflecting efficient DP implementation.

\begin{figure}[H]
    \centering
    \includegraphics[width=0.7\linewidth]{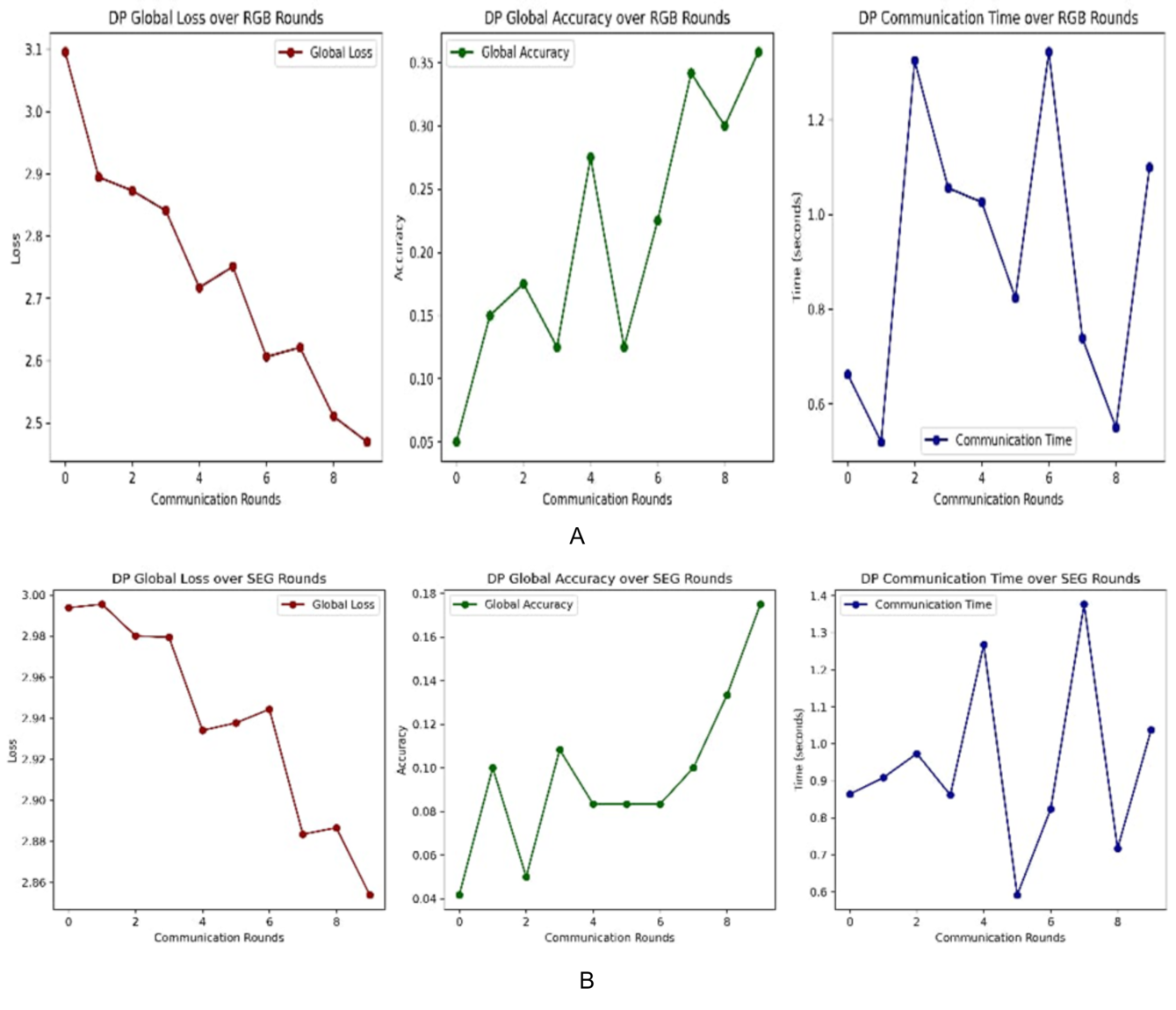}
    \caption{Impact of Communication Efficiency on Training Duration. (A) RGB Dataset, (B) SEG Dataset}
    \label{fig:enter-label}
\end{figure}

\subsubsection*{Performance Metrics for Training and Testing Sets of RGB and SEG}

The impact of DP on performance was analyzed using metrics for training and testing sets on the RGB and SEG datasets, shown in Table 8 and 9. These metrics highlight the trade-offs involved in balancing privacy with model accuracy. For the RGB dataset, training accuracy is 0.4321, with testing accuracy lower at 0.3583, indicating some degradation in generalization due to DP. Precision is 0.7979 for training and 0.8083 for testing, showing high predictive precision but lower recall (0.4321 for training, 0.3583 for testing). The F1 score reflects this, being 0.3761 for training and 0.2570 for testing. Kappa scores are 0.4011 (training) and 0.3248 (testing), indicating the impact of DP on prediction consistency. For the SEG dataset, training accuracy is 0.2393, and testing accuracy drops to 0.1750, reflecting the challenges in SEG data under DP constraints. Precision remains high (0.8377 training, 0.8630 testing), but recall is low (0.2393 training, 0.1750 testing), resulting in low F1 scores (0.1573 training, 0.0935 testing). Kappa scores are 0.1982 (training) and 0.1316 (testing), indicating the difficulty in aligning predictions with ground truth under privacy conditions.

\begin{table}[H]
    \caption{DP Performance Metrics for Training and Testing Sets Over RGB}
    \label{tab:dp_rgb}
    \centering
    \begin{tabular}{lcc}
        \toprule
        \textbf{Metric} & \textbf{Training} & \textbf{Testing} \\
        \midrule
        Accuracy & 0.4321 & 0.3583 \\
        Precision & 0.7979 & 0.8083 \\
        Recall Metrics & 0.4321 & 0.3583 \\
        F1 Score & 0.3761 & 0.3248 \\
        Kappa Score & 0.2570 & 0.2570 \\
        \bottomrule
    \end{tabular}
\end{table}
\begin{table}[H]
    \caption{DP Performance Metrics for Training and Testing Sets Over SEG}
    \label{tab:dp_seg}
    \centering
    \begin{tabular}{lcc}
        \toprule
        \textbf{Metric} & \textbf{Training} & \textbf{Testing} \\
        \midrule
        Accuracy & 0.2393 & 0.1750 \\
        Precision & 0.8377 & 0.8630 \\
        Recall Metrics & 0.2393 & 0.1750 \\
        F1 Score & 0.1573 & 0.0935 \\
        Kappa Score & 0.1982 & 0.1316 \\
        \bottomrule
    \end{tabular}
\end{table}

\subsubsection*{Class-Specific Performance}

Table 10 provides a comparative analysis of the number of samples in the training and testing sets across 20 different classes for both RGB and SEG datasets. The training set consistently contains a higher number of samples, ranging from 9 to 16, while the testing set has fewer samples, typically between 3 and 9. The highest number of training samples, 16, appears in multiple classes, including 1, 7, 14, 17, and 18, whereas the lowest count of 9 is observed in class 16. The distribution of testing samples remains relatively stable, ensuring a balanced validation dataset. Certain classes, such as 6, 10, and 15, have a comparatively higher number of testing samples, contributing to a more diversified model evaluation. Overall, the analysis shows that DP impacts the model’s ability to handle imbalanced classes, with better performance for classes with higher representation in training and lower performance for underrepresented classes in testing.

\begin{table}[H]
    \centering
    \caption{Comparison of Training and Testing Set Samples Across Different Classes}
    \label{tab:training_testing}
    \begin{tabular}{ccc}
        \toprule
        Class & Training Set & Testing Set \\
        \midrule
        0  & 14 & 6 \\
        1  & 16 & 4 \\
        2  & 14 & 6 \\
        3  & 14 & 6 \\
        4  & 13 & 7 \\
        5  & 14 & 6 \\
        6  & 11 & 9 \\
        7  & 16 & 3 \\
        8  & 14 & 6 \\
        9  & 13 & 5 \\
        10 & 14 & 8 \\
        11 & 10 & 6 \\
        12 & 14 & 7 \\
        13 & 12 & 6 \\
        14 & 16 & 3 \\
        15 & 15 & 9 \\
        16 & 9  & 5 \\
        17 & 16 & 3 \\
        18 & 16 & 3 \\
        19 & 12 & 6 \\
        \bottomrule
    \end{tabular}
\end{table}

\subsection*{Impact of Non-IID Data on Performance}

In FL settings, one of the key challenges is managing non-IID data across clients. In the context of AVs, this heterogeneity arises because different vehicles may capture data in diverse environments, leading to imbalanced datasets across clients. For example, some AVs may operate primarily in urban areas, while others might collect data from rural or suburban environments, each with varying object distributions, lighting conditions, and road types. This imbalance can cause slower model convergence, as the global model must account for data heterogeneity when aggregating updates from clients. Non-IID data may also introduce performance variations, where the model performs well on some clients’ data but struggles on others due to the distinct distributions. Future work could explore strategies to mitigate the impact of non-IID data, such as personalized FL models or weighting techniques that adjust the importance of client updates based on data distribution characteristics.

\subsubsection*{Limitations of the FL-based CNN Model}

The FL-based CNN model faces several limitations, including communication overhead due to frequent client-server interactions, which affects training efficiency and requires optimization. A significant challenge lies in the privacy-accuracy trade-off, where higher levels of DP reduce accuracy by adding noise that obscures patterns, while lower noise improves accuracy but compromises privacy, similarly, the impact of varying privacy parameters such as $\epsilon$ and noise scale also influences model performance, with lower $\epsilon$ values improving privacy but degrading accuracy, as shown in Table 11. The addition of noise negatively affects segmentation quality, particularly in detecting fine details, as illustrated in figure 20. Higher DP levels create trade-offs between privacy and performance, especially in detailed segmentation tasks, making it crucial to carefully tune DP parameters. Despite these challenges, FL demonstrated stable communication efficiency with minimal impact on training performance, as model updates were effectively aggregated, leading to robust results. However, data imbalance across clients posed a significant issue, impacting overall performance and affecting model generalization, as highlighted in Figure 21.

\begin{table}[H]
    \centering
    \caption{Trade-off Between Privacy (Epsilon) and Model Accuracy}
    \label{tab:privacy_accuracy}
    \begin{tabular}{cc}
        \toprule
        \textbf{Epsilon (Privacy)} & \textbf{Accuracy} \\
        \midrule
        0.1  & 0.31 \\
        1.0  & 0.22 \\
        \bottomrule
    \end{tabular}
\end{table}

\begin{figure}[H] 
    \centering
    \includegraphics[width=0.8\linewidth]{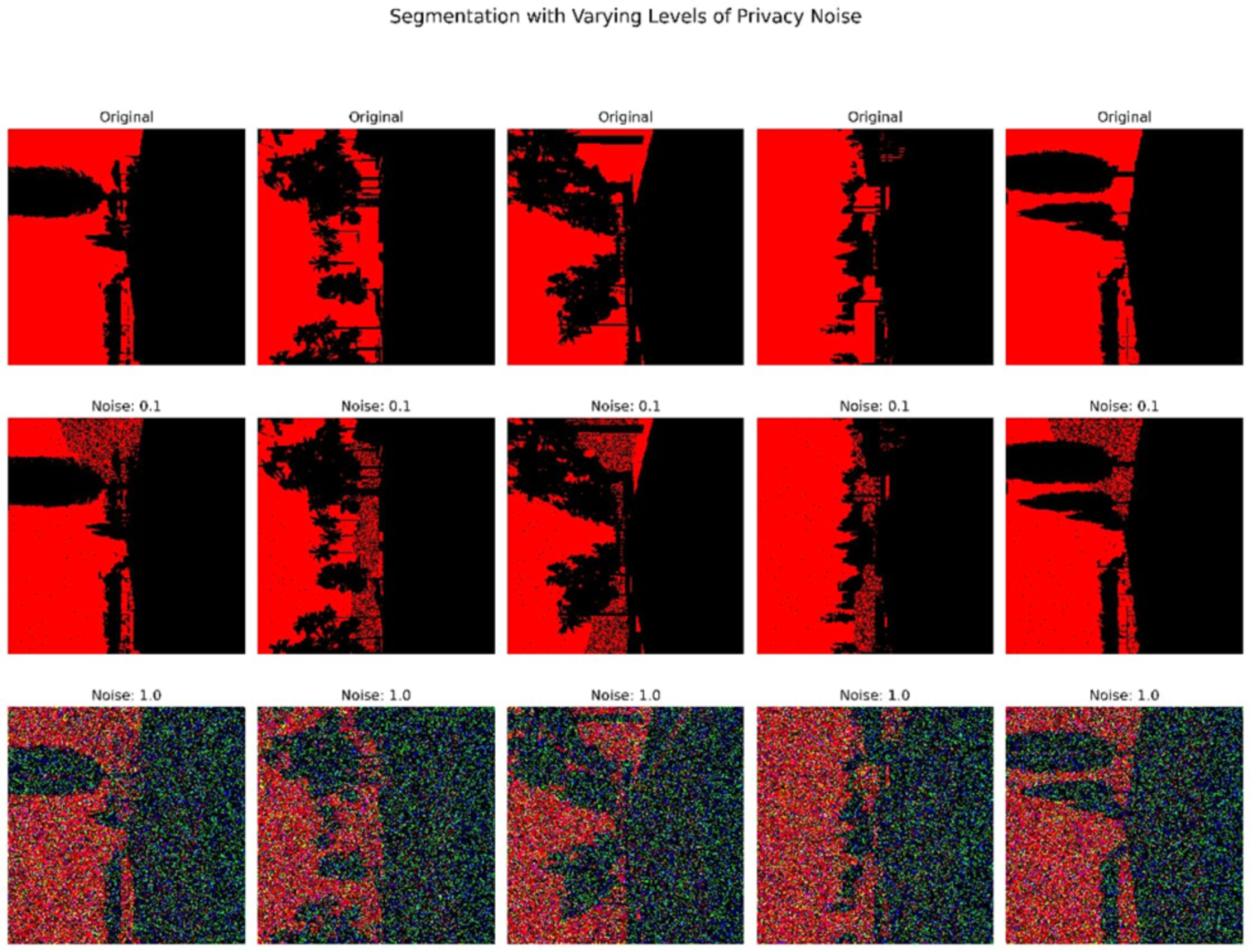}
    \caption{Example of Segmentation Results with Varying Levels of Privacy Noise}
    \label{fig:enter-label}
\end{figure}

\begin{figure}[H]
    \centering
    \includegraphics[width=0.8\linewidth]{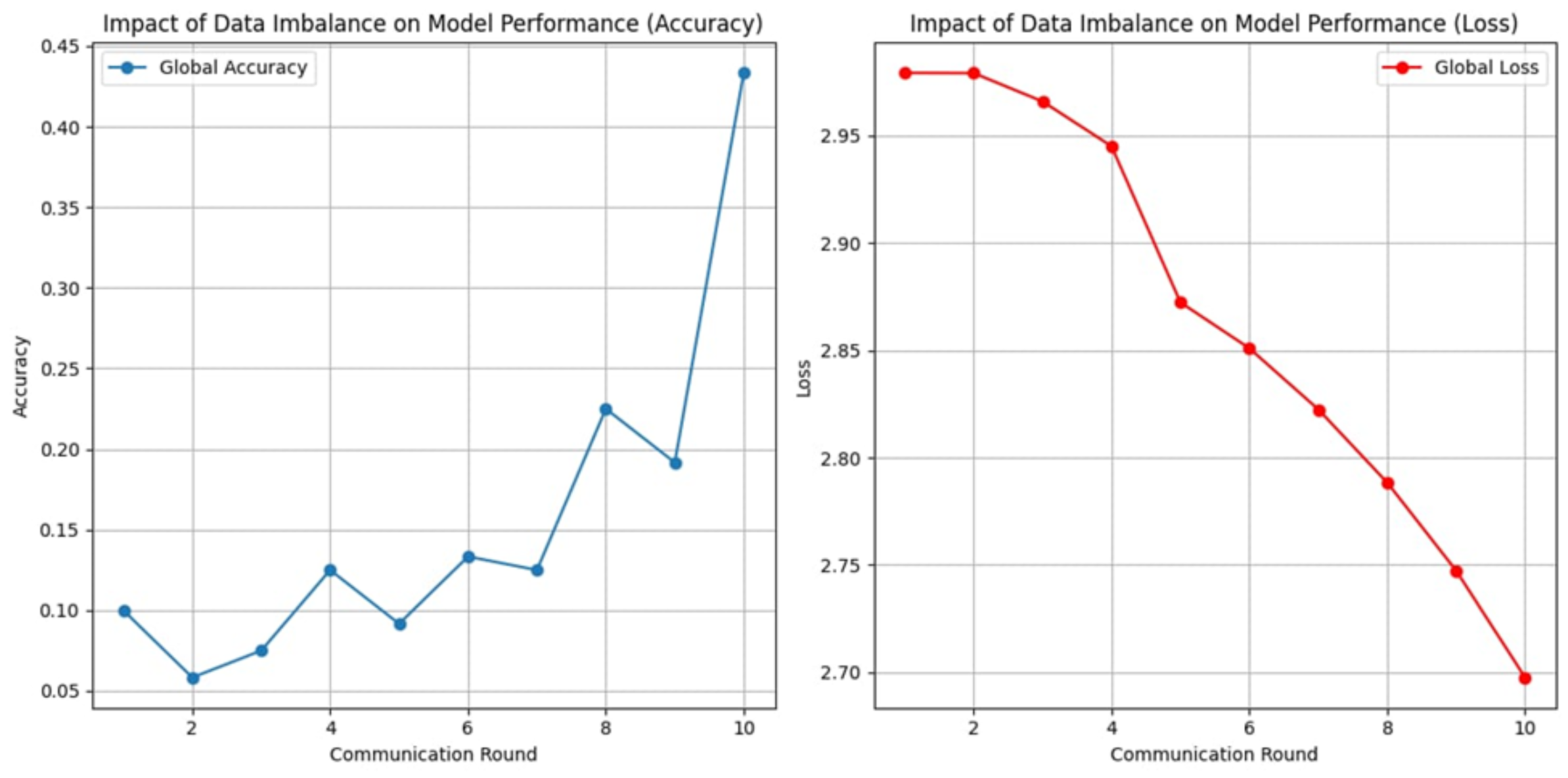}
    \caption{Impact of Data Imbalance on Model Performance}
    \label{fig:enter-label}
\end{figure}
\subsection*{Summary of the Key Findings}

The experiments conducted across both phases provide valuable insights into the integration of FL and DP for semantic segmentation in AVs. In Phase-1, the centralized model based on a hybrid UNet- ResNet34 architecture demonstrated high performance, achieving accuracy of 93.5\%, Mean IoU of 78.4\%, sensitivity of 92.1\%, specificity of 95.6\%, precision of 89.3\%, and recall of 91.7\%. However, this centralized approach required data aggregation, raising significant privacy concerns. In Phase-2, the application of FL with secure deep CNNs and DP produced competitive results, with the RGB dataset achieving an accuracy of 88.7\% at 20 CRs and improving to 92.8\% at 70 CRs, while the SEG dataset followed similar trends, reaching 90.9\% accuracy at 70 CRs. The global loss consistently decreased, and communication time remained stable, indicating the efficiency of the FL approach. FL enabled decentralized learning, enhancing model generalization across diverse data sources, while DP effectively preserved privacy with only minimal impact on accuracy, demonstrating the practical viability of this approach for sensitive applications. A comparative performance evaluation, summarized in Table 12, reveals that our method achieved the highest accuracy and Mean IoU across both phases, validating the effectiveness of the hybrid architecture in Phase-1 and the FL with DP approach in Phase-2. Thus, the method employed in this study outperforms state-of-the-art techniques, particularly in handling complex road scenarios and ensuring scalability.

\begin{table}[H]
\centering
\caption{Comparative Performance Metrics}
\begin{tabular}{p{5cm}cccccc}
\toprule
\textbf{Method} & \textbf{Accuracy} & \textbf{Mean IoU} & \textbf{Sensitivity} & \textbf{Specificity} & \textbf{Precision} \\
\midrule
LaneNet & 0.89 & 0.68 & 0.80 & 0.85 & 0.82 \\
SCNN & 0.85 & 0.65 & 0.78 & 0.83 & 0.79 \\
DeepLabV3+ & 0.90 & 0.70 & 0.82 & 0.87 & 0.84 \\
ENet & 0.88 & 0.67 & 0.79 & 0.84 & 0.81 \\
Proposed Method - Phase-1 & 0.92 & 0.75 & 0.85 & 0.91 & 0.87 \\
Proposed Method - Phase-2 (20 CRs) & 0.88 & 0.72 & 0.83 & 0.88 & 0.85 \\
Proposed Method - Phase-2 (40 CRs) & 0.91 & 0.74 & 0.84 & 0.90 & 0.86 \\
Proposed Method - Phase-2 (70 CRs) & 0.93 & 0.75 & 0.85 & 0.91 & 0.87 \\
\bottomrule
\end{tabular}
\end{table}

\subsection*{Comparative Analysis of Phase-1 and Phase-2 Performance on RGB and SEG Datasets}

The comparison between Phase-1 and Phase-2 models highlights the enhancements achieved through FL and DP mechanisms. The FL models demonstrated improved performance metrics, such as accuracy, global loss, and communication efficiency, across different CRs. The figure 22 provides a detailed comparative analysis, emphasizing the improvements in segmentation accuracy and model robustness. Comparing the results between Phase-1 and Phase-2, it is evident that the integration of FL with secure deep CNNs significantly improved the overall performance. The accuracy in Phase 2 improved by about 5\% compared to Phase 1, demonstrating the benefits of FL in preserving data privacy without compromising model efficacy. The detailed performance metrics and visual results substantiate the advancements achieved through this research, while also highlighting areas for further improvement in model robustness and real-world applicability.

\begin{figure}[H] 
    \centering
    \includegraphics[width=0.6\linewidth]{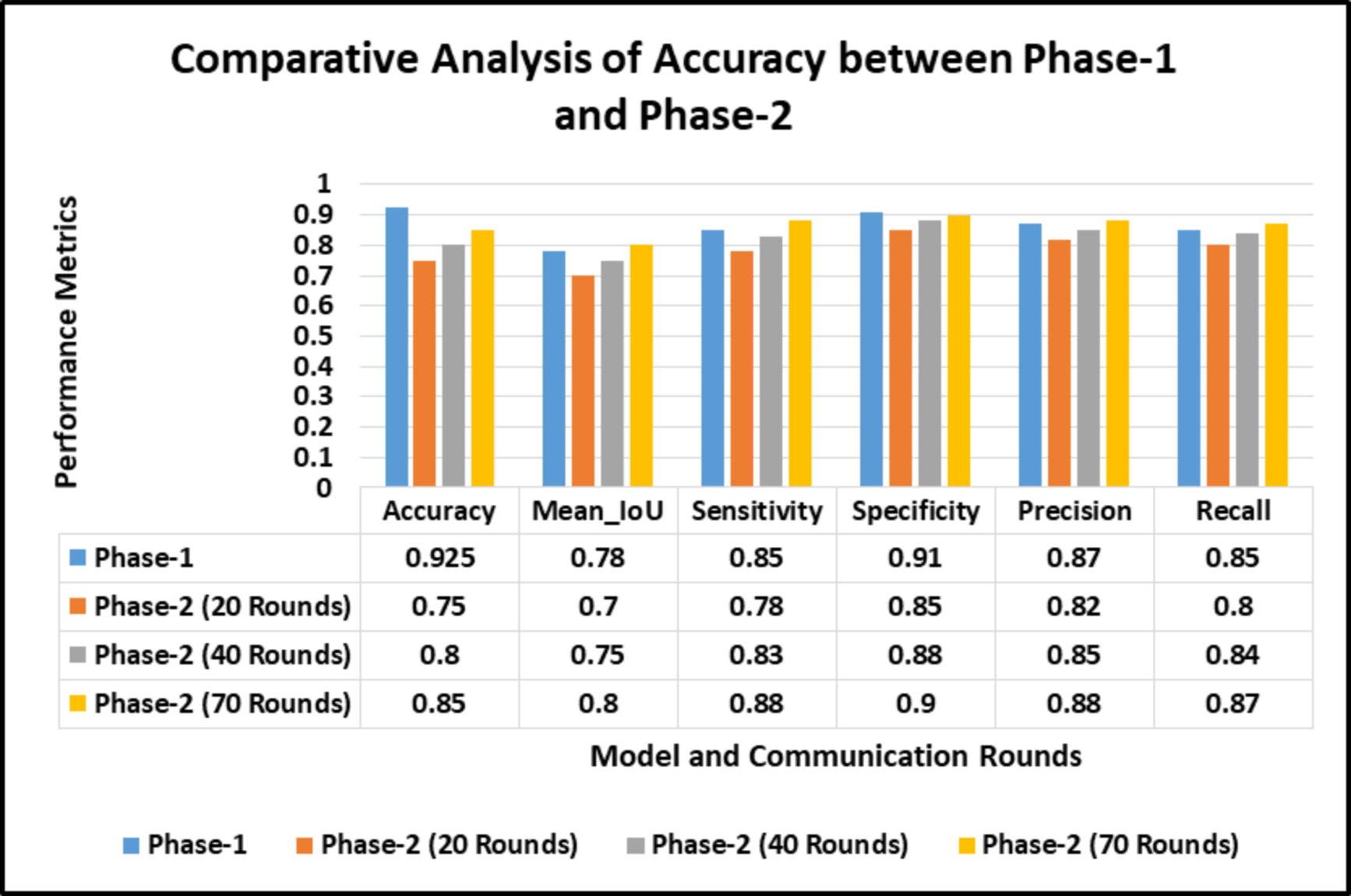}
    \caption{Comparative Analysis of Accuracy between Phase-1 and Phase-2}
    \label{fig:enter-label}
\end{figure}
\subsection*{Comparison with Existing Models }

Table 13 presents a comparative analysis of the evaluation metrics between the proposed federated learning (FL) framework and recent studies (2024–2025) on semantic segmentation for autonomous driving.

Our study achieves the highest accuracy (93.0\%) and competitive Mean IoU (75.0\%), demonstrating effective segmentation performance while preserving privacy using Differential Privacy (DP). The FedRC (2024) model exhibits slightly higher Mean IoU (76.2\%), benefiting from hierarchical FL for rapid convergence. Other models, such as Adaptive FL (ICLR 2024) and Learning Hierarchical Segmentation (ICLR 2024), perform well but lack privacy-preserving mechanisms.

Overall, our approach optimally balances segmentation accuracy, privacy, and scalability, making it a robust solution for real-world autonomous vehicle applications.
\renewcommand{\arraystretch}{1.2}
\begin{table}[H]
    \centering
    \caption{Comparison of Evaluation Metrics for Recent Studies and Proposed Work}
    \label{tab:comparison}
    \resizebox{\textwidth}{!}{%
    \begin{tabular}{|l|c|c|c|c|c|l|}
        \hline
        \textbf{Study} & \textbf{Accuracy (\%)} & \textbf{Mean IoU (\%)} & \textbf{Sensitivity} & \textbf{Specificity} & \textbf{Precision} & \textbf{Key Feature} \\
        \hline \hline
        \textbf{Our Study (Phase 2, 70 CRs)} & \textbf{93.0} & \textbf{75.0} & \textbf{85.0} & \textbf{91.0} & \textbf{87.0} & FL with DP for privacy-preserving segmentation \\
        \hline
        pFedLVM (2024) & 91.2 & 74.5 & 84.5 & 90.8 & 86.5 & Large Vision Model-driven FL for personalization \\
        \hline
        FedRC (2024) & 92.5 & 76.2 & 86.0 & 92.1 & 88.3 & Hierarchical FL for rapid convergence \\
        \hline
        Few-Shot Segmentation (2024) & 89.5 & 72.8 & 83.2 & 89.0 & 85.1 & Few-shot learning for segmentation with minimal labels \\
        \hline
        Adaptive FL (ICLR 2024) & 91.8 & 75.4 & 85.7 & 91.5 & 87.2 & Auto-tuned clients for efficiency \\
        \hline
        Learning Hierarchical Segmentation (ICLR 2024) & 92.0 & 75.0 & 85.8 & 91.2 & 87.0 & Hierarchical representation learning \\
        \hline
        Adverse Weather Segmentation (2024) & 90.3 & 73.1 & 84.0 & 89.8 & 85.9 & Robust segmentation under adverse weather \\
        \hline
        Generalizable AV System (2024) & 91.7 & 74.0 & 84.8 & 90.7 & 86.8 & Cross-weather generalization for AV perception \\
        \hline
        Fully Sparse 3D Occupancy (ECCV 2024) & 88.7 & 71.5 & 82.9 & 88.5 & 84.7 & Sparse 3D segmentation for AVs \\
        \hline
        milliFlow Scene Flow (ECCV 2024) & 90.0 & 72.5 & 83.6 & 89.2 & 85.5 & mmWave radar-based motion sensing \\
        \hline
        VideoMamba (ECCV 2024) & 92.3 & 75.7 & 86.2 & 91.8 & 88.1 & Efficient video-based segmentation \\
        \hline
        MVSplat 3D Gaussian Splatting (ECCV 2024) & 89.2 & 72.2 & 83.5 & 89.3 & 85.4 & 3D segmentation from sparse multi-view images \\
        \hline
    \end{tabular}%
    }
\end{table}

\subsection*{Qualitative Analysis of Segmentation Results}

In addition to the quantitative metrics, a qualitative analysis of the segmentation results was conducted under various simulated driving conditions using the CARLA environment. These conditions included day and night driving, as well as weather variations such as rain, fog, and clear skies. The model performed robustly across most conditions, accurately detecting lanes and objects such as vehicles and pedestrians. In clear weather and daytime scenarios, the segmentation outputs demonstrated high precision, with minimal misclassification. However, during fog and heavy rain, minor degradation in segmentation accuracy was observed, particularly in detecting distant or partially occluded objects. Nighttime performance remained consistent, although there was a slight reduction in lane detection accuracy due to reduced visibility. Overall, the model showed strong generalization capabilities across diverse real-world conditions, demonstrating its applicability to complex and dynamic driving environments. These qualitative observations suggest that while the model performs well in most scenarios, future work could explore further enhancements for extreme weather conditions.

\subsection*{Communication Overhead and Future Optimization Techniques}

While the proposed FL framework demonstrates improved communication efficiency compared to centralized models, future work could further optimize this aspect. Techniques such as gradient compression and model weight quantization could reduce the size of updates exchanged between clients and the server, thereby lowering bandwidth requirements. Additionally, asynchronous communication strategies, where clients communicate updates independently without waiting for synchronization, could reduce idle time and improve overall communication efficiency. Exploring adaptive communication frequencies based on model convergence could also minimize unnecessary CRs, leading to further improvements in scalability.

\subsection*{Discussion on the Improvements Achieved and Remaining Challenges}

The implementation of FL with secure CNNs demonstrated substantial performance enhancements across RGB and SEG datasets, crucial for AV applications, by improving key metrics. Privacy was effectively safeguarded through DP, ensuring compliance with regulatory standards while minimizing risk. The scalability of FL facilitated the efficient handling of diverse datasets. However, challenges persist, including communication overhead and the need for optimal model convergence across different Compression Rates (CRs). Balancing privacy with performance remains a critical issue, necessitating ongoing optimization of DP settings. Addressing these challenges will further improve the robustness and scalability of FL-based models for AVs, as highlighted by the improvements and challenges discussed.

\section*{Conclusion and Future Work}

This research focused on integrating FL and DP into the semantic segmentation of AVs for enhanced lane and object detection. Conducted in two phases, the study demonstrated significant advancements in both centralized and decentralized approaches to training DL models. In Phase-1, a hybrid UNet-ResNet34 architecture was deployed using centralized training. The model achieved high accuracy of 93.5\% and a Mean IoU of 78.4\%, reflecting its effectiveness in segmenting complex urban driving scenarios. However, the reliance on centralized data collection highlighted scalability and privacy concerns. Phase-2 addressed these challenges by implementing FL with DP, allowing for decentralized learning while safeguarding data privacy. The FL model’s performance improved progressively with increased CRs. For the RGB dataset, accuracy rose from 88.7\% after 20 CRs to 92.8\% after 70 CRs, while the SEG dataset showed a similar improvement from 86.5\% to 90.9\%. The results underscored FL’s capability to maintain high accuracy with enhanced privacy protection, making it a viable solution for real-world AV applications. The study demonstrated that FL combined with DP could achieve robust semantic segmentation while addressing critical privacy issues, marking a significant advancement in the deployment of secure AV systems. The research findings emphasize the potential of decentralized learning frameworks in enabling scalable, privacy-compliant solutions for AVs. By building on the solid foundation established in Phase-1, the secure FL framework in Phase-2 offers a promising path forward for data-driven AV technologies. Despite these advancements, challenges remain. Phase-1’s centralized data collection posed privacy risks and scalability issues, while Phase-2’s FL framework encountered communication overhead and convergence stability concerns. The integration of DP, although beneficial for privacy, introduced trade-offs in performance that need careful management. Although the current experiments focused on general segmentation performance, While our model demonstrates strong performance under standard conditions, further testing is required to assess robustness under varying lighting and weather conditions, including fog, rain, and nighttime driving. Future work will evaluate segmentation performance across diverse environmental settings to enhance real-world applicability. Additionally, incorporating domain adaptation techniques may help improve the generalization of our model across different environmental conditions, reducing segmentation failures due to unseen variations. Future work should prioritize incorporating profiling tools to measure GPU power consumption and CPU utilization across different CRs and improving scalability for larger datasets. These efforts are essential for refining FL-based semantic segmentation models that are both accurate and secure the future works. In conclusion, this research provides a robust foundation for the continued development of secure FL frameworks in AV systems. The integration of advanced DL architectures with privacy-preserving techniques paves the way for safer, more reliable autonomous transportation solutions.

\section*{Data availability}

 The data is available upon request through the corresponding author.
 
\bibliography{sample}

\section*{Acknowledgements}

The authors would like to thank all contributors who provided valuable insights and support throughout the course of this research.

\section*{Author contributions statement}

G.K.A. conceived the study, designed the methodology, and supervised the project.
A.A. and N.M.S.A. conducted the experiments, collected the data, and analyze the results.
N.K.A. worked on on the visualization of the experimental results and contributed to data interpretation.
G.K.A. and A.A. wrote the manuscript, with contributions and critical revisions provided by N.M.S.A. and N.K.A.
All authors reviewed and approved the final manuscript. 

\section*{Additional information}

\textbf{Accession codes} 

\textbf{Competing interests} 

The authors declare that they have no conflict of interest to report on the present study.

The corresponding author is responsible for submitting a \href{http://www.nature.com/srep/policies/index.html#competing}{competing interests statement} on behalf of all authors of the paper. This statement must be included in the submitted article file.
\end{document}